# Electrocaloric cooling cycles in lead scandium tantalate with true regeneration via field variation

S. Crossley[1,+], B. Nair[1], R. W. Whatmore[2], X. Moya[1] and N. D. Mathur[1,*]

[1]Materials Science, University of Cambridge, Cambridge, CB3 0FS, *United Kingdom*

[2]*Department of Materials, Royal School of Mines, South Kensington Campus, Imperial College London, London SW7 2AZ, United Kingdom*

[*]Corresponding author: ndm12@cam.ac.uk

[+]Currently at the Department of Applied Physics, Stanford University, Stanford, CA 94305, USA

There is growing interest in heat pumps based on materials that show thermal changes when phase transitions are driven by changes of electric, magnetic or stress field. Importantly, regeneration permits sinks and loads to be thermally separated by many times the changes of temperature that can arise in the materials themselves. However, performance and parameterization are compromised by net heat transfer between caloric working bodies and heat-transfer fluids. Here we show that this net transfer can be avoided—resulting in true, balanced regeneration—if one varies the applied electric field while an electrocaloric (EC) working body dumps heat on traversing a passive fluid regenerator. Our EC working body is represented by bulk $PbSc_{0.5}Ta_{0.5}O_3$ (PST) near its first-order ferroelectric phase transition, where we record directly measured adiabatic temperature changes of up to 2.2 K. Indirectly measured adiabatic temperature changes of similar magnitude were identified, unlike normal, from adiabatic measurements of polarization, at nearby starting temperatures, without assuming a constant heat capacity. The resulting high-resolution field-temperature-entropy maps of our material, and a small clamped companion sample, were used to construct cooling cycles that assume the use of an ideal passive regenerator in order to span ≤20 K. These cooling cycles possess well-defined coefficients of performance that are bounded by well-defined Carnot limits, resulting in large (>50%) well-defined efficiencies that are not unduly compromised by a small field hysteresis. Our approach permits the limiting performance of any caloric material in a passive regenerator to be established, optimized and compared; provides a recipe for true regeneration in prototype cooling devices; and could be extended to balance active regeneration.

Cooling technology based on vapour-compression is widespread, but its reliance on environmentally harmful gases has led to growing interest in prototype heat pumps[1-17] that exploit EC, magnetocaloric (MC) and mechanocaloric (mC) materials. In these materials[18,19], phase transitions—or other transitions such as those seen in relaxors[6,20,21]—are driven by changes of electric, magnetic and stress field, respectively. By exploiting passive[1-4,6,9,11] or active[4,7,12,13,15] regeneration, caloric materials can pump heat across temperature spans that greatly exceed the magnitude of the field-driven adiabatic temperature change $\Delta T$ in any given caloric material. In the case of passive regeneration, a nominally homogeneous caloric working body establishes a large temperature differential along a much longer column of fluid that constitutes the regenerator. In the case of active regeneration, a relatively small volume of heat-transfer fluid is used to establish a large temperature differential along both itself and a bed of caloric material(s) that constitute the regenerator. However, cooling cycles that have been hitherto proposed or realized do not display true (balanced) regeneration because the zero-field and finite-field isofields for a material are not parallel, forcing a deleterious net transfer of heat between caloric material and regenerator. In the case of passive regeneration, this net transfer has been noted explicitly in a seminal paper[1], and elsewhere[22].

The key figure of merit for any cooling device is its coefficient of performance COP = $Q/W$, where in each cooling cycle, work $W$ is done to pump heat $Q$ from a cold load at $T_c$ towards a hot sink at $T_h$. The effect of the net-transfer problem is to partially subsume regenerators into sinks or loads, with two consequences. First and foremost, it becomes impossible to realize intended cooling cycles, thus compromising device COPs. Second, the active material undergoes net heat exchange at temperatures lying away from $T_c$ and $T_h$, thus compromising the calculation of the Carnot limit $T_c/(T_h - T_c)$ from measured values of $T_c$ and $T_h$. Given that COPs and Carnot limits are thus compromised, it immediately follows that device efficiencies are neither optimized nor accurate (efficiency = COP/Carnot). To solve the net-transfer problem, we will construct valid cooling cycles on detailed field-temperature-entropy maps, and we will accurately evaluate the resulting COPs, Carnot limits and efficiencies. These parameters contrast with COPs for materials in isothermal cycles[23], and the related parameter of materials efficiency[24] that describes reversible caloric transitions driven isothermally in just one direction.

The net-transfer problem has been hitherto masked in EC cooling cycles because the zero-field and finite-field isofields are rendered parallel by assuming peak performance at all points in any given cycle[25-27]. However, even for relaxors that show large EC effects over wide ranges of temperatures[6,20,21], this assumption is only reasonable for small temperature spans $T_h - T_c$ that would only be useful when exploiting converse EC effects for pyroelectric energy harvesting[28]. The net-transfer problem has also been masked in cooling cycles based on mC materials, either by likewise assuming peak performance[29] or by assuming isothermal conditions[23]. In the larger body of work on MC cooling, the net-transfer problem has arisen even after identifying zero-field and maximum-field isofields[1,22,30-32], reflecting the need for field variation during regenerator transit.

Three solutions to the net-transfer problem have been hitherto proposed for MC materials, but they seem far from perfect. First, one could engineer a composite MC material that displays a large and constant isothermal entropy change over the target temperature span[33], but this would be laborious and require a large number of component materials to approach perfection. Second, one could exploit serendipity under very specific circumstances, as shown using entropy-temperature data that were obtained for intermediate fields using a mean-field model[34], but these circumstances are too restrictive to be of any practical value. Third, one could hold sample magnetization constant during regenerator transit[35], but this is experimentally unrealistic (and does not account for the small variation of sample entropy with temperature).

In this paper, we demonstrate a general solution to the net-transfer problem by constructing cooling cycles in which no net heat is exchanged between a homogeneous working body of the archetypal EC material PST[2,18,36-39] and an ideal hypothetical regenerator that it traverses in order to achieve a large temperature span $T_h - T_c \gg |\Delta T|$. The underlying principle is that the two regenerator-transit legs of a given cooling cycle can be made to differ by a constant entropy if the field applied during the finite-field leg is varied according to a detailed $E(T,S')$ map of the highly reversible phase transition at finite fields above Curie temperature $T_C \sim 295$ K, where $E$ denotes electric field, $T$ denotes temperature, and $S'$ denotes entropy $S$ after subtracting the zero-field entropy at our base temperature of 285 K. In contrast with the highly restrictive solutions discussed above[33-35], our strategy for true regeneration via field variation can be readily achieved by modifying standard cooling cycles, such as Ericsson cycles. Here we modify Brayton cycles because they involve adiabatic EC effects that can be driven quickly, thus increasing cooling power and reducing heat leaks.

The $E(T,S')$ maps are detailed and accurate enough to identify the field variation required for true regeneration because they are derived by implementing the well known[18] indirect (Maxwell) method with a range of improvements that we explain in Methods. The two most notable improvements are as follows. First, our electrical polarization data are acquired at measurement set temperatures whose separation is two orders of magnitude less than the ~10 K separation of standard practice[20,40]. Second, we evaluate adiabatic temperature change versus starting temperature without assuming some constant value of the specific heat capacity[18]. Our improved implementation of the indirect method results in detailed maps of $|\Delta T(S',E)|$ that we corroborate by measuring temperature change directly with a thermocouple. Adding the starting temperature yields maps of absolute temperature $T(S',E)$, and permuting the variables yields maps of the entropy state function $S'(T,E)$. The resulting maps of $|\Delta S(T,E)|$ show a strong temperature dependence that is responsible for the regeneration imbalance between cycle legs that follow zero-field and finite-field isofields. The resulting maps of $E(T,S')$ permit the construction of cooling cycles with true regeneration via field variation. For all of the afore-mentioned colour maps, we present selected cross-sections in Supplementary Information, permitting numerical values to be read with ease.

Our cycle efficiencies are found to exceed ~50% even for our largest temperature span of $T_h - T_c = 20$ K, and even after accounting for the small field hysteresis of the transition. Similar efficiencies are also obtained for a small volume of PST. This sample was clamped by its unelectroded surroundings, and could withstand larger electric fields. Consequently it displayed a wider temperature range of operation, and per unit volume of PST would be able to pump more heat from some load in each cycle. Our PST cycle efficiencies represent accurate upper bounds on the efficiencies of EC prototypes were they to be based on our material, and large magnitude of these efficiencies reveals that EC cooling is an intrincisally efficient process. In future, our approach should permit accurate comparison of caloric materials assuming the use of ideal regenerators, and improve the parameterization and performance of regenerative cooling cycles.

**Results**

PST was fabricated as described in Methods. Dielectric measurements confirmed the expected first-order ferroelectric phase transition at $T_C \sim 295$ K, with a small hysteresis of 2.6 K, and low dielectric loss (Fig. 1a). Increasing the measurement frequency from 100 Hz to 100 kHz increased the temperature of the dielectric peak by just ~1 K (Supplementary Fig. 1), consistent with the suppression of the relaxor behaviour[37] that is expected with our high degree of B-site cation order (~0.80).

The first-order phase transition is also seen in the specific heat capacity $c(T)$ measured on heating (Fig. 1b), from which we evaluate entropy $S'(T) = S(T) - S(285\text{ K})$ on heating (Fig. 1b), as explained in Methods. The entropy change for the full transition $|\Delta S_0| \sim 28.4$ kJ K$^{-1}$ m$^{-3}$ (see Methods) corresponds to the difference between $S'(T)$ above $T_C$ and the background value with no transition (Fig. 1b). As expected for a predominantly displacive phase transition in the solid state, $|\Delta S_0|$ is small with respect to the thermally induced changes of entropy away from the transition. These thermally induced changes are subtracted in cooling cycles that employ a regenerator[1], but the subtraction is only precise if one is able to achieve balanced regeneration.

We will now use the indirect method to quantify changes of temperature that arose during our adiabatic (constant $S'$) measurements of electrical polarization $P(E)$ with field step

~0.3 kV cm$^{-1}$, and a maximum field of magnitude $E_{max}$ = 26 kV cm$^{-1}$. The resulting bipolar and unipolar plots were obtained at 252 measurement set temperatures $T(E = 0)$ that we identified every ~0.16 K on heating slowly through the ferroelectric transition at 0.5 K min$^{-1}$. Four representative bipolar (Fig. 1c) and unipolar (Fig. 1d) plots form part of the full dataset that appears in Supplementary Fig. 2. The four bipolar plots show that the ferroelectric loop at 280 K became a double loop[41] over a limited range of temperatures above $T_C$, evidencing the field-driven transition in PST (ref. 36). However, we will use our unipolar plots in what follows because the field hysteresis is smaller. Temperature and entropy will be interconverted at zero field from the heating branch of $S'(T)$ (Fig. 1b), such that an electric field will modify $S'(T)$ without entering the thermally hysteretic region.

For each sign of field, the field-application branches of our unipolar $P(E)$ plots at 252 values of $S'$ were transposed to yield $P(S')$ plots at 100 values of $E$, after identifying these 100 values of $E$ by linearly interpolating between the ~83 experimental values of $E$ that differed slightly in each measured branch. The 100 plots of $P(S')$ were smoothed using cubic spline fits, and are presented as $|P(S',E)|$ (Fig. 2a) with the measurement set temperature $T(E = 0)$ marked on a secondary abscissa using $S'(T)$ (Fig. 1b). The plot of $|P(S',E)|$ represents a first-order phase diagram, and the resulting map of $|\partial P/\partial S|_E$ (Fig. 2b) highlights the phase boundary along which no critical point was reached. By exploiting the Maxwell relation $(\partial T/\partial E)_S = -(\partial P/\partial S)_E$ that is relevant for adiabatic measurements of polarization (Supplementary Note 3), this latter map permits calculation of the nominally reversible adiabatic temperature change $\Delta T(S',E) = -\int_0^E (\partial P/\partial S)_{E'} dE' > 0$ with respect to starting temperature $T(E = 0)$ due to the application of field $E$ (Fig. 2c). The two 'teeth' are seen to bite at a value of $S'$ that corresponds to $T_C$ ~ 295 K, and there is a threshold field for discernible EC effects at higher starting temperatures. The addition of $\Delta T(S',E)$ to starting temperature $T(S',0)$ (identified from $S'(T)$ in Fig. 1b) results in a map of absolute temperature $T(S',E)$ (Fig. 2d). While following an isothermal contour on this map, one may read the isothermal entropy change $\Delta S'(T,E)$ from the entropy abscissa. The temperature for any such contour may be read at zero field via the secondary abscissa $T(E = 0)$.

The panels of Fig. 2e-h were constructed in an analogous manner to the panels of Fig. 2a-d, using the field-removal branches of our unipolar $P(E)$ data, such that the starting temperature $T(E = \pm E_{max})$ shown on the secondary abscissa exceeds the measurement set temperature $T(E = 0)$ due to the adiabatic EC temperature change that occurred during each field-application branch (Fig. 2c,d). The nominally reversible adiabatic temperature change $\Delta T(S',E) = -\int_{\pm E_{max}}^{E} (\partial P/\partial S)_{E'} dE' < 0$ was mapped for $\pm E_{max} \to E$ (Fig. 2g), such that it was largest for small $E$ and thus complementary with respect to the 'teeth' of Fig. 2c. One might have instead expected $\Delta T(S',E)$ to be evaluated for $E \to 0$, but in view of the small field hysteresis, this would more properly require field-removal data for multiple values of $E_{max}$ at each measurement temperature. In practice, the field hysteresis is too small to significantly compromise our single-valued assumption on $P(S',E)$, such that the resulting maps of absolute temperature $T(S',E)$ are similar when derived from field-application and field-removal branches (Fig. 2d,h).

Direct measurements of EC temperature change in PST were performed using a thermocouple (see Methods), and the resulting figure panels (Fig. 2i-p) are ordered for ready comparison with their indirect counterparts (Fig. 2a-h), rather than in the order that they were obtained. At a series of measurement set temperatures $T(E = 0)$ that differed by 0.5 K, we repeated the following well-known Brayton cycle with applied fields that were increased by 1.1 kV cm$^{-1}$ for the next cycle. An adiabatic temperature increase $\Delta T > 0$ due to field application ($0 \rightarrow E$) was followed by a slow isofield return to the measurement set temperature $T(E = 0)$, and a subsequent adiabatic temperature decrease $\Delta T < 0$ due to field removal ($E \rightarrow 0$) was likewise followed by a slow isofield return to the same measurement set temperature. All of the resulting time-dependent data appear in Supplementary Fig. 3. After converting measurement set temperature to entropy via $S'(T)$ (Fig. 1b), we plotted $\Delta T(S',E)$ (Fig. 2k,o) and $T(S',E)$ (Fig. 2l,p) for field application (Fig. 2k,l) and field removal (Fig. 2o,p). Our directly measured EC effects (Fig. 2l,p) are similar for field application and field removal, and similar to our indirectly measured EC effects (Fig. 2d,h). A more detailed comparison regarding the magnitude of these EC effects appears in Supplementary Note 7.

The directly measured maps of temperature change $|\Delta T(S',E)|$ (Fig. 2k,o) may be used to back out $|\partial P/\partial S|_E$ (Fig. 2j,n) and thus $|P(S',E)|$ (Fig. 2i,m), assuming at each field an integration constant of $P \sim \varepsilon_0 \varepsilon E$ at $S' = 350$ kJ K$^{-1}$ m$^{-1}$, where $\varepsilon = 5730$ (Fig. 1a) and $\varepsilon_0$ is the permittivity of free space. Electrical data of the type used to deduce EC effects via the indirect method have therefore themselves been deduced via directly measured EC effects, with the caveat that this method cannot identify the remanent polarization of the ferroelectric phase.

Permuting the variables in all four $T(S',E)$ maps (Fig. 2d,h,l,p) resulted in four entropy maps $S'(T,E)$ (Fig. 3a,e,i,m). While following an adiabatic contour on one of these entropy maps, one may read the adiabatic temperature change $\Delta T(S',E)$ from the temperature abscissa. The entropy for any such contour may be read at zero field via the secondary abscissa $S'(E = 0)$. As expected, the ambipolar phase boundary $|dE/dT| \sim 1$ kV cm$^{-1}$ K$^{-1}$ is correctly predicted by the Clausius-Clapeyron equation $|dE/dT| = |\Delta S_0|/|\Delta P|$, given a thermally driven entropy change of $|\Delta S_0| \sim 28.4$ kJ K$^{-1}$ m$^{-3}$ (Fig. 1b) and spontaneous polarisation of $|\Delta P| \sim 26$ μC cm$^{-2}$ at 280 K (Fig. 1d).

The four entropy maps (Fig. 3a,e,i,m) were used to map the magnitude of the reversible entropy change $|\Delta S(E,T)|$ (Fig. 3b,f,j,n) that would arise from the isothermal application or removal of field $E$ at temperature $T$. The strong temperature dependence of $|\Delta S(E,T)|$ can be eliminated by varying the field, and employing this field variation during the regenerator transit at finite field yields balanced regeneration. In a small range of temperatures just above $T_C$, the field-driven entropy change $|\Delta S(E,T)|$ at sufficiently high fields roughly corresponds to the zero field entropy change for the thermally driven transition ($|\Delta S_0| \sim 28.4$ kJ K$^{-1}$ m$^{-3}$). This imples that the electrically driven transition has been driven to completion, which is rare in the EC literature, and results in a reduced isothermal EC strength $|\Delta S(E,T)|/|E|$ at our highest fields (Fig. 3c,g,k,o). The persistence of a large entropy change $|\Delta S(E,T)|$ well above $T_C$ is important because it implies that large temperature spans $T_h$ - $T_c$ could be achieved in regenerative devices based on PST alone. Our entropy maps permit us to identify the isofield specific heat capacity $c(E,T) = T(\partial S/\partial T)_E$ (Fig. 3d,h,l,p), which successfully captures the low-field sharpness[42] of the peak in $c(T)$, and demonstrates an equivalence between our implementation of the indirect method and the quasi-direct method[18], where caloric effects are evaluated by measuring the field and temperature dependence of the heat capacity.

On permuting the variables in our first entropy map $S'(T,E)$ (Fig. 3a), which shows more conservative EC effects than its counterpart derived from field-removal data (Fig. 3b), we identified $E(T,S')$ (Fig. 4a) for positive fields above $T_C$ in order to construct Brayton-like cooling cycles with true regeneration. For a range of load and sink temperatures ($T_c$ and $T_h$), these Brayton-like cycles will be parameterized while initially neglecting the small field hysteresis of the transition, resulting in COPs that depend purely on the choice of cycle in accessible parameter space (these COPs will be denoted COP$_{cyc}$). We will then account for the small field hysteresis of the transition to obtain COPs that represent upper bounds for cooling devices which are based on real PST but otherwise ideal (these COPs will be denoted $COP_{mat}$).

A Brayton-like cooling cycle (1→2→X→3→4→Y→1) with specific values of $T_h$ and $T_c$ is shown in Fig. 4a. Most of the electrical work is done (1→2) and recovered (3→4) when driving EC effects adiabatically. Heat is dumped under isofield conditions (2→X) to a sink at $T_h$, and absorbed at zero field (4→Y) from a load at $T_c$. The intervening steps involve an idealised regenerator (inset, Fig. 4a), which heats PST from $T_c$ to $T_h$ during transit in zero field (Y→1), and cools PST from $T_h$ to $T_c$ during transit in finite field (X→3).

This finite field would assume the constant value of $E_{max}$ if one were to follow the standard practice that results in a net transfer of heat between the EC working body and the regenerator. Here we ensure that the regenerator and PST exchange the same heat (and entropy) in each half cycle by translating Y→1 down the entropy axis in order to identify X→3, such that these two legs differ by a constant entropy at all temperatures. By translating Y→1 as far down the entropy axis as possible, X→3 intersects the $E_{max}$ isofield, thus maximizing the heat pumped from the load under the constraint of true regeneration.

The field variation required to follow X→3 (Fig. 4a) may be seen more clearly when the cycle is presented on (*T,E*) axes (Fig. 4b), where 1→2 and 3→4 are adiabatic contours such as those in Fig. 3a. Presenting the cycle on (*E,P*) axes (red data and green isofields, Fig. 4f) reveals that *P* is fairly constant during finite-field regenerator transit X→3. Given that *P* ~ 0 during the zero-field regenerator transit Y→1, the difference of polarization between the two regenerator legs is fairly constant, as expected given that the constant entropy difference has been achieved for a range of temperatures that is too small to substantially modify phonon populations. One could therefore approximate true regeneration by maintaining constant polarization during X→3, which would in practice mean electrical isolation to achieve constant charge density $D \sim P$. However, it is experimentally straightforward to vary the applied field, and we seek to identify the true upper bounds on $COP_{mat}$, so we will follow the precise recipe for true regeneration.

By varying $T_h$ and $T_c$ for the Brayton-like cycles 1→2→X→3→4→Y→1 on our map of $E(T,S')$ (Fig. 4a), we plot on axes of $(T_c, T_h - T_c)$ the heat $Q = \int_4^Y T(S') \, dS'$ absorbed from the load (Fig. 4c) as calculated from the zero-field specific heat capacity data via $S'(T)$ (Fig. 1b); the cycle work $W = \oint T dS'$ (Fig. 4d) as calculated from the area of the cycle on $(T,S')$ axes; and thus $COP_{cyc} = Q/\oint T dS'$ (Fig. 4e). Our use of reversible thermodynamics is justified because all heat exchange with respect to the PST working body should be slow given the finite thermal time constant of the sink (2→X), the finite thermal time constant of the load (4→Y), and the need to avoid undue turbulence and friction during regenerator transit (Y→1 and X→3).

We may equally evaluate $COP_{cyc} = Q/\oint E dP$ via the cycle area on $(E,P)$ axes (red data and green isofields, Fig. 4f), where electrical work $\oint E dD \sim \oint E dP$. We are then able to crudely account for the small field hysteresis of the transition (Fig. 1d) purely by recalculating the work done to drive a given cycle whose area is increased by replacing the field-removal leg (red X→4, Fig. 4f) derived from the field-application branches of $|P(S',E)|$ (Fig. 2a) with a field-removal leg (blue X→4, Fig. 4f) derived from the field-removal branches of $|P(S',E)|$ (Fig. 2e). The resulting values of $W = \oint E đP$ (Fig. 4g) slightly exceed the values of $W = \oint T dS' = \oint E dP$ (Fig. 4d) that were calculated without taking the field hysteresis into account (đP denotes an inexact differential). Realistic values of $COP_{mat} = Q/\oint E đP$ (Fig. 4h) are therefore slightly smaller than anhysteretic values of $COP_{cyc} = Q/\oint T dS'$ (Fig. 4e).

For useful temperature spans of $T_h - T_c = 10$ K and 20 K, cross-sections through our map of $Q$ (Fig. 4c) are presented in Fig. 5a, while the corresponding cross-sections through our maps of $COP_{cyc}$ (Fig. 4e) and $COP_{mat}$ (Fig. 4h) are presented after dividing through by the Carnot limit to yield cycle efficiency (Fig. 5b). A simple geometric construction shows that our values of $COP_{cyc}$ only fall short of 100% because of triangular areas 1–2–X and 3–4–Y, which render the EC effects near $T_h$ and $T_c$ adiabatic in order to avoid the impractically slow isothermal EC effects that would arise in balanced Ericsson-like cycles 1→X→3→Y→1. Importantly for applications, our realistic values of $COP_{mat}$ exceed ~50% of the Carnot limit, suggesting that PST is a promising material for regenerative cooling devices.

To see how the embodiment of an EC material can influence $COP_{mat}$, we also investigated a clamped sample of PST, whose broadened phase transition was driven by order-of-magnitude larger electric fields (Supplementary Notes 11-15). Our clamped PST permitted a larger volume-normalized heat $Q$ to be achieved across a much wider temperature range (Fig. 5c), without significantly compromising efficiency (Fig. 5d). The EC properties of PST and clamped PST are compared with other EC materials in Supplementary Note 16.

COPs for a PST-based cooling device would necessarily be smaller than the values of $COP_{mat}$ that we report here. This is because of engineering losses such as those associated with real regenerators, those associated with the imperfect recovery[17,24] of electrical work during depolarization (2→4), and those associated with the inactive margins of EC working bodies. However, it is encouraging to find that our values of $COP_{mat}$ for both PST and clamped PST (12-27 for $T_h - T_c = 10$ K, 8-12 for $T_h - T_c = 20$ K) comfortably exceed the COPs for a state-of-the-art cooling device[43] based on MC gadolinium (3 for $T_h - T_c = 10$ K, zero for $T_h - T_c = 20$ K). Moreover, it would be easy to vary the electric field during X→3 for true regeneration, and it would be easy to maintain constant charge for nearly true regeneration. By contrast, it could be challenging to vary the applied magnetic field from permanent magnets for true regeneration in MC cycles, and it would presumably be even more challenging to maintain constant magnetization for nearly true regeneration[35].

**Discussion**

In this paper, we have obtained indirect and direct measurements of the EC effects that arise in ~400 μm-thick polycrystalline $PbSc_{0.5}Ta_{0.5}O_3$ when the ferroelectric phase transition is electrically driven over a wide range of starting temperatures that lie above the Curie temperature, and we have found that this range can be extended to considerably higher temperatures by clamping. Our indirect and direct measurements were obtained by collecting data at nearby temperatures and fields, resulting in detailed maps that interrelate these variables via entropy. The subsequent construction of Brayton-like cooling cycles above $T_C$ produced three key findings. First, one may achieve true regeneration by varying the applied field when a homogeneous caloric material dumps heat on traversing an ideal fluid regenerator, such that the resulting cooling cycles are formally valid with well-defined COPs, Carnot limits and efficiencies. Second, hysteretic losses associated with the field-driven transition are small compared with the work done to pump heat, such that our regeneratively balanced cycles are highly efficient. Third, the demonstration of large efficiencies in balanced cooling cycles demonstrates that our EC material is promising for applications.

Our work has three practical implications for the future. First, all types of caloric material should be evaluated, compared and selected via analyses of the type that we show here for PST. To recap, this involves collecting dense data, producing detailed maps of field on axes of entropy and temperature, choosing a range of load and sink temperatures, constructing a given type of cooling cycle with true regeneration via field variation, taking field hysteresis into account as necessary, and calculating well-defined COPs and efficiencies. Second, prototype heat pump COPs should be improved by realizing true regeneration. This would involve first using detailed maps of the active caloric material in order to identify the field variation required for true regeneration with an ideal regenerator, and then refining the field variation (e.g. via machine learning) to account for engineering losses such as those we described earlier. Third, realizing true regeneration in prototype heat pumps would permit device efficiencies (COP/Carnot) to be calculated with precision. Ultimately, detailed thermodynamic maps such as those we present here could be used to identify the field variation required to balance any type of cooling cycle, including the cooling cycles used for active regeneration.

**Acknowledgements.** We thank P. C. Osbond for fabricating the master sample, O. Idigoras for help with sample preparation, and A. P. Carter, E. Defay, R. J. Harrison, I. Takeuchi, A. Rowe and J. Tušek for discussions. X. M. is grateful for support from the Royal Society. B. N. is grateful for support from Gates Cambridge and the Winton Programme for the Physics of Sustainability. We thank EPSRC (UK) for funding via EP/M003752/1 and a DTA award (S. C.).

**Methods**

**Sample preparation.** Samples A-E were derived from a 0.79 mm-thick master wafer of polycrystalline PbSc$_{0.5}$Ta$_{0.5}$O$_3$ (PST), which was fabricated using the mixed-oxide method described in ref. 44. Specifically, Sc$_2$O$_3$ and Ta$_2$O$_5$ powders were milled together and then prereacted at 900 °C to form the wolframite phase ScTaO$_4$. This phase was then reacted with PbO at 900 °C to form a single-phase perovskite powder, which was subsequently hot-pressed in Si$_3$N$_4$ tooling and an alumina grit packing medium at 40 MPa and 1200 °C for 6 hours. The B-site cation order was inferred[45,46] to be ~0.80 from $|\Delta S_0| \sim 28.4$ kJ K$^{-1}$ m$^{-3}$, and we assume a density[36] of $\rho \sim 9071$ kg m$^{-3}$.

Samples A-D were produced by thinning, electroding and mounting as described below, with more details in ref. 47, where a schematic of the entire procedure appears on page 108. Sample E comprised 13.249 mg of unthinned and unelectroded PST, derived from the master wafer.

Samples A-C are described as 'PST'. Using P1200 grit polishing paper and isopropanol, these samples were fabricated by hand-thinning pieces of the master wafer to a thickness of ~400 μm, such that Sample A was 420 μm thick, Sample B was 390 μm thick, and Sample C was 450 μm thick. The concomitant reduction of applied voltage permitted arcing to be avoided in our measurement probe. Sample D is described as 'clamped PST'. In order to increase the field that could be applied without breakdown, it was fabricated by hand-thinning a piece of the master wafer to 95 μm. P4000 paper was used to polish the side on which the bottom electrode was deposited, and P1200 and then P4000 paper were used to polish the other side.

Pt electrodes of thickness ~70 nm were deposited on Samples A-D by sputtering. The bottom electrode covered the entire lower surface of each sample, while the smaller top electrode fell short of sample edges to avoid arcing. This shortfall was ~0.5 mm for Samples A-C, and several millimetres for Sample D. Nominal top-electrode areas of 32 mm$^2$ (Sample A), 39 mm$^2$ (Sample B), and 30 mm$^2$ (Sample C) include a ~6% increase for fringing fields[48]. For Sample D, we avoided arcing by applying Apiezon 'N' grease to the small top electrode, whose effective area was ~0.42 mm$^2$ (Supplementary Note 11).

Samples A-D were mounted on an electrically insulated Cu substrate using a thin layer of silver paint (Samples A-C) or silver epoxy (Sample D).

**Calorimetry.** This was performed by using a TA Instruments Q2000 to heat Sample E at 5 K min$^{-1}$. At temperature $T$, the change of entropy with respect to the absolute entropy at 285 K was evaluated via $S'(T) = S(T) - S(285\,\text{K}) = \int_{285\text{K}}^{T} c(T')/T'\,dT'$ (Fig. 1b). The entropy change across the full transition $|\Delta S_0| = \left|\int_{T_1}^{T_2} c(T')/T'\,dT'\right| \sim 28.4$ kJ K$^{-1}$ m$^{-3}$ was obtained after subtracting the baseline background, with $T_1$ set freely below the transition, and $T_2$ set freely above the transition.

**Electrical measurements.** These were performed using a cryogenic probe that was fabricated in house[47]. Dielectric measurements were performed using an Agilent 4294A analyser at 0.1-100 kHz, while ramping the temperature at ±1 K min$^{-1}$. Electrical polarization was measured on heating, using a Radiant Precision Premier II with a Trek high-voltage amplifier. Similar data were obtained on cooling[47]. All unipolar plots were positioned on the polarization axis by centring the corresponding bipolar plots.

The electrical polarization data for PST were collected under adiabatic conditions, given that the 0.25 s measurement timescale (insets of Fig. 1c-d) is 20 times smaller than the ~5 s timescale for heat transfer that we identify from thermocouple measurements using the same cryostat (Supplementary Fig. 3).

The electrical polarization data for clamped PST were also collected under adiabatic conditions. This follows because the 0.025 s measurement timescale (insets of Supplementary Fig. 10a,b) was 10 times smaller than the 0.25 s measurement timescale for PST, while the thermal time constant was reduced by a factor of ~20 with respect to PST, such that the resulting measurement timescale was 10 times smaller than the resulting thermal time constant. (Thermal time constant varies quadratically with sample thickness, as thermal resistance and thermal capacitance each vary linearly with sample thickness.)

**Improvements to the indirect (Maxwell) method.** The first of two key improvements was to acquire $P(E)$ data dense in temperature, thus permitting the construction of detailed $E(T,S')$ maps on which we could identify the field variation required for true regeneration. Our second key improvement was to quantify the adiabatic temperature change as a function of starting temperature without following the standard practice[18] of assuming a constant specific heat capacity $c$. Instead, we generated maps of $c(T,E)$ (Fig. 3) from our $P(E)$ and $c(T)$ data.

Three other improvements to the indirect method were as follows. (1) We employed a short measurement timescale in order to obtain adiabatic rather than isothermal $P(E)$ data, thus reducing the time required for data collection. As a consequence, the adiabatic temperature change was evaluated via Maxwell relation $(\partial T/\partial E)_S = -(\partial P/\partial S)_E$ (Supplementary Note 3) without knowledge of the specific heat capacity, which was nevertheless required in order to convert entropy to starting temperature. (2) Our unipolar plots of $P(E)$ permitted us to minimise the possibility of ferroelectric domain switching at the small subset of measurement temperatures lying near and below $T_C$. Moreover, unipolar cycles would likely reduce fatigue under service conditions. (3) We obtained similar EC effects using the field-application and field-removal branches of our unipolar $P(E)$ data. This simultaeneously confirmed that the single-valued assumption on $P(S',E)$ is reasonable, and that the field-driven first-order phase transition is highly reversible above $T_C$.

**Direct measurements of temperature change in PST.** These were performed using a spot-welded K-type thermocouple, whose active junction was attached with silver paste to Sample C at a location lying near the centre of the top electrode. The two wires that represent the reference junction were each attached separately with silver paste to an insulating layer of Kapton tape, which covered the surface of the heat reservoir in the cryogenic probe that we used for electrical measurements. Starting well above $T_C \sim 295$ K, data were obtained at measurement set temperatures every ~0.5 K from 320 K down to 285 K. The direction of this temperature sweep was unimportant given that the large EC effects of interest lie above $T_C$ and are therefore nominally reversible (Fig. 2k,o).

Using a Keithley 2410 sourcemeter, we applied and removed electric fields on millisecond timescales in order to drive highly adiabatic changes of temperature. However, our measured jumps in thermoelectric voltage $V$ did not capture these temperature changes in full, for two reasons. First, the temperature of the active volume was necessarily modified by heat exchange with the thermocouple. Second, heat exchange between the active volume and the thermocouple was in practice accompanied by exchange heat between the active volume and the inactive material (the unaddressed PST, the sample mounting, and the very thin top electrode).

To scale our measured temperature jumps into the adiabatic limit, we established a calibration factor of $\Delta T/\Delta V = 62.97$ K mV$^{-1}$ for the thermocouple by removing 15 kV cm$^{-1}$ from PST at ambient temperature, and equating the $\Delta V = 22$ μV response of Sample C with the temperature jump of $\Delta T \sim 1.4$ K that we measured for Sample B using scanning thermal microscopy (SThM, Supplementary Fig. 4a). Our calibration factor exceeds the calibration factor for K-type thermocouples by a factor of 2.5 for the two reasons discussed above, and is used to present all data arising from direct measurements, namely the measured temperature changes themselves [$T(t)$, Supplementary Fig. 3], the temperature jumps that we extracted from these measurements [$\Delta T(S',E)$, Fig. 2k,o], and the resulting plots of absolute temperature [$T(S',E)$, Fig. 2l,p].

The aforementioned SThM measurements were performed at a tip-sample separation of 20 μm, using an Anasys head attached to a Veeco Multimode atomic force microscope, and a Keithley 3116 voltmeter to monitor the temperature signal at a sampling rate of kHz. The EC temperature change measured by SThM is assumed to be adiabatic because the heat capacity of the sample was seven orders of magnitude larger than the ~1 nJ K$^{-1}$ heat capacity of the tip. The accuracy of the SThM calibration was limited by both the ~0.2 K noise in the SThM data, and a small mismatch of absolute temperature that may have occurred with respect to the thermocouple data under calibration.

An infra-red point-detector (CS LT 15, Optris GmbH) was used to measure EC effects at ambient temperature as described in ref. 47, and the results for Sample A (Supplementary Fig. 4b) were approximately consistent with our SThM measurements of Sample B.

**Cross-sections through colour maps.** These represent a permutation of the variables for selected values of the ordinate. For example, cross-sections through a colour map of $S'(T,E)$ (Fig. 3a) are presented as $E(T,S')$ (Supplementary Fig. 6a) for three values of $E$. Repeating this procedure for all values of $E$ yields the full map of $E(T,S')$ (Fig. 4a).

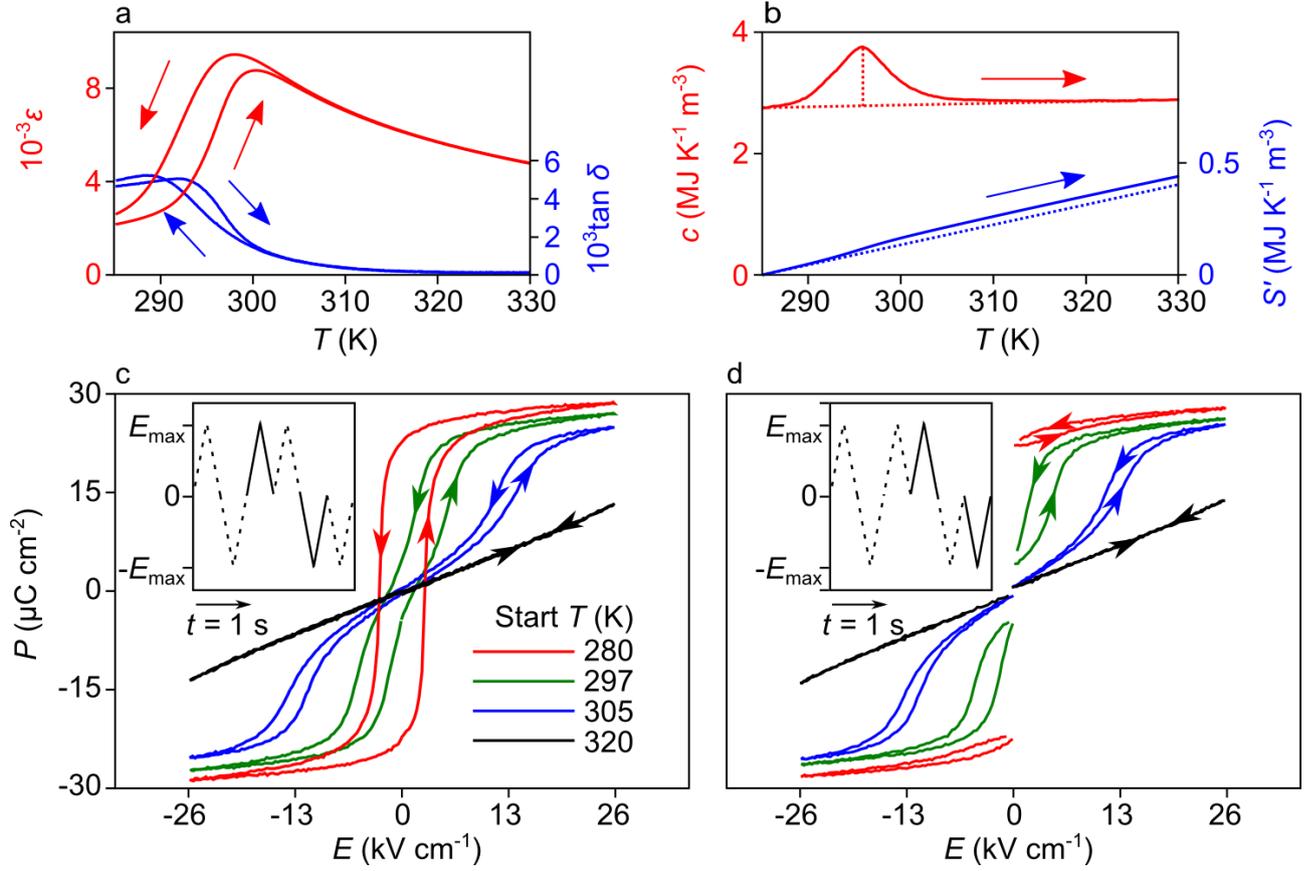

**Figure 1. PST characterization and measurement.** (a,b) The zero-field phase transition. (a) Relative dielectric constant $\varepsilon$ and loss tangent tan $\delta$ versus temperature $T$. (b) Specific heat capacity $c(T)$ (red), dotted lines denote baseline and peak. Hence entropy $S'(T) = S(T) - S(285\text{ K})$ (blue), dotted line denotes linear extrapolation from below $T_C \sim 295$ K. (c,d) Adiabatic polarization $P(E)$ with $E_{max} = 26$ kV cm$^{-1}$, for selected starting temperatures on heating. Data for (c) bipolar and (d) unipolar plots were acquired during the times denoted by solid lines on the insets, which show driving field $E$ versus time $t$. Supplementary Fig. 1 shows the frequency dependence of $\varepsilon$. Supplementary Fig. 2 shows bipolar and unipolar $P(E)$ plots for all 252 measurement temperatures. Data in (a,c,d) for Sample A. Data in (b) for Sample E.

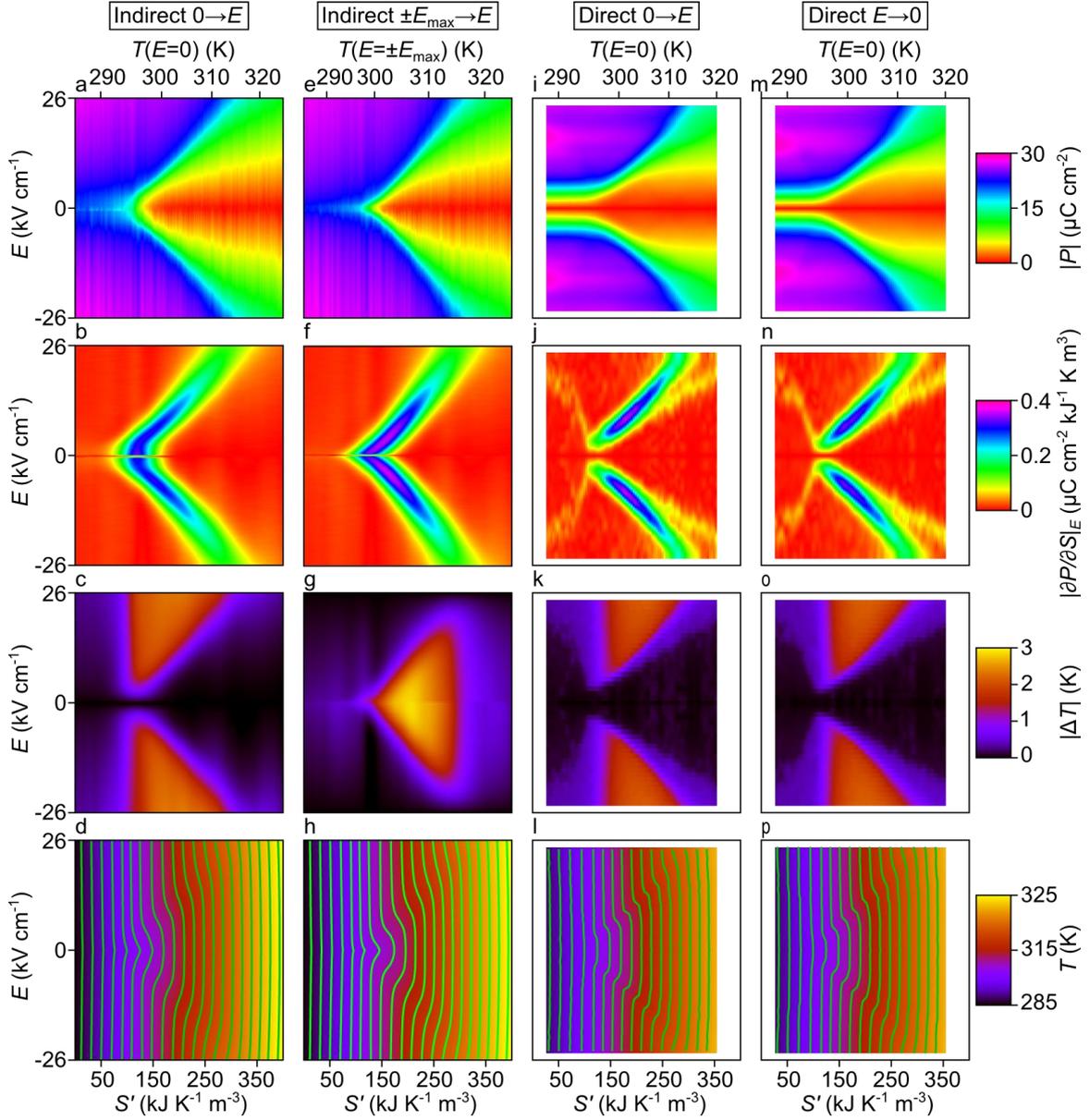

**Figure 2. PST polarization and temperature on axes of entropy and field.** (a-h) Data based on indirect EC measurements for (a-d) $0 \to E$ and (e-h) $\pm E_{max} \to E$. (a,e) For each sign of field, $|P(S',E)|$ was constructed by plotting 100 isofield cubic spline fits $P(S')$ to (a) field-application branches ($0 \to E$) and (e) field-removal branches ($\pm E_{max} \to E$) of 252 unipolar $P(E)$ plots (Supplementary Fig. 2) obtained at starting temperatures separated by 0.16 K. The resulting plots of (b,f) $|\partial P/\partial S|_E$ imply nominally reversible adiabatic temperature changes of (c,g) $|\Delta T(S',E)|$ starting at (c) $T(E=0)$ and (g) $T(E=\pm E_{max})$. Hence (d,h) $T(S',E)$. (i-p) Data based on direct EC measurements for (i-l) field application ($0 \to E$) and (m-p) field removal ($E \to 0$). We identified (j,n) $|\partial P/\partial S|_E$ and (i,m) $|P(S',E)|$ from (k,o) the directly measured temperature change $|\Delta T(S',E)|$ with respect to equilibrium temperature $T(E=0)$, which we also used to identify (l,p) $T(S',E)$. All direct measurements of temperature change (Supplementary Fig. 3) were obtained for $E > 0$, such that all data in (i-p) are mirrored at negative field. (d,h,l,p) Isothermal contours every ~1.7 K represent every third measurement set temperature in (l,p). Data in (a-h) for Sample A with $E_{max} = 26$ kV cm$^{-1}$. Data in (i-p) for Sample C with $E_{max} = 24.5$ kV cm$^{-1}$. Variation of starting temperature $T(E=0)$ (top abscissa) with entropy $S'$ (bottom abscissa) from $S'(T)$ (Fig. 1b). Constant-field cross-sections appear in Supplementary Fig. 5.

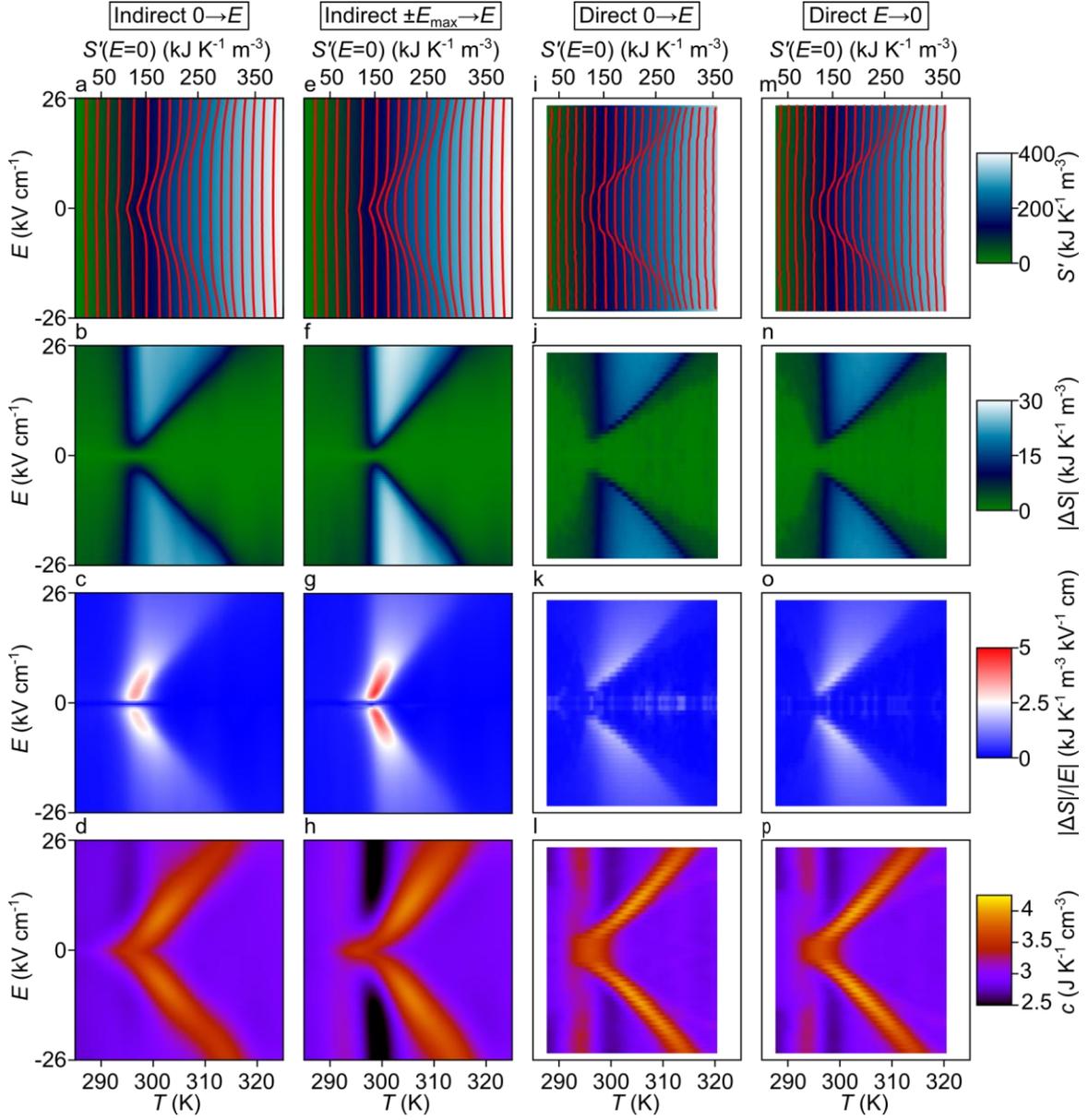

**Figure 3. PST entropy and specific heat capacity on axes of temperature and field.** (a,e,i,m) Entropy $S'(T,E)$ obtained by permuting the variables in $T(S',E)$ (Fig. 2d,h,l,p), with adiabatic contours every ~20 kJ K$^{-1}$ m$^{-3}$. Hence (b,f,j,n) the nominally reversible isothermal entropy change $|\Delta S(T,E)|$ for $0 \leftrightarrow E$ at temperature $T$. (c,g,k,o) EC strength $|\Delta S(T,E)|/|E|$. (d,h,l,p) Specific heat capacity $c(T,E) = T(\partial S/\partial T)_E$. Constant-field cross-sections appear in Supplementary Fig. 6.

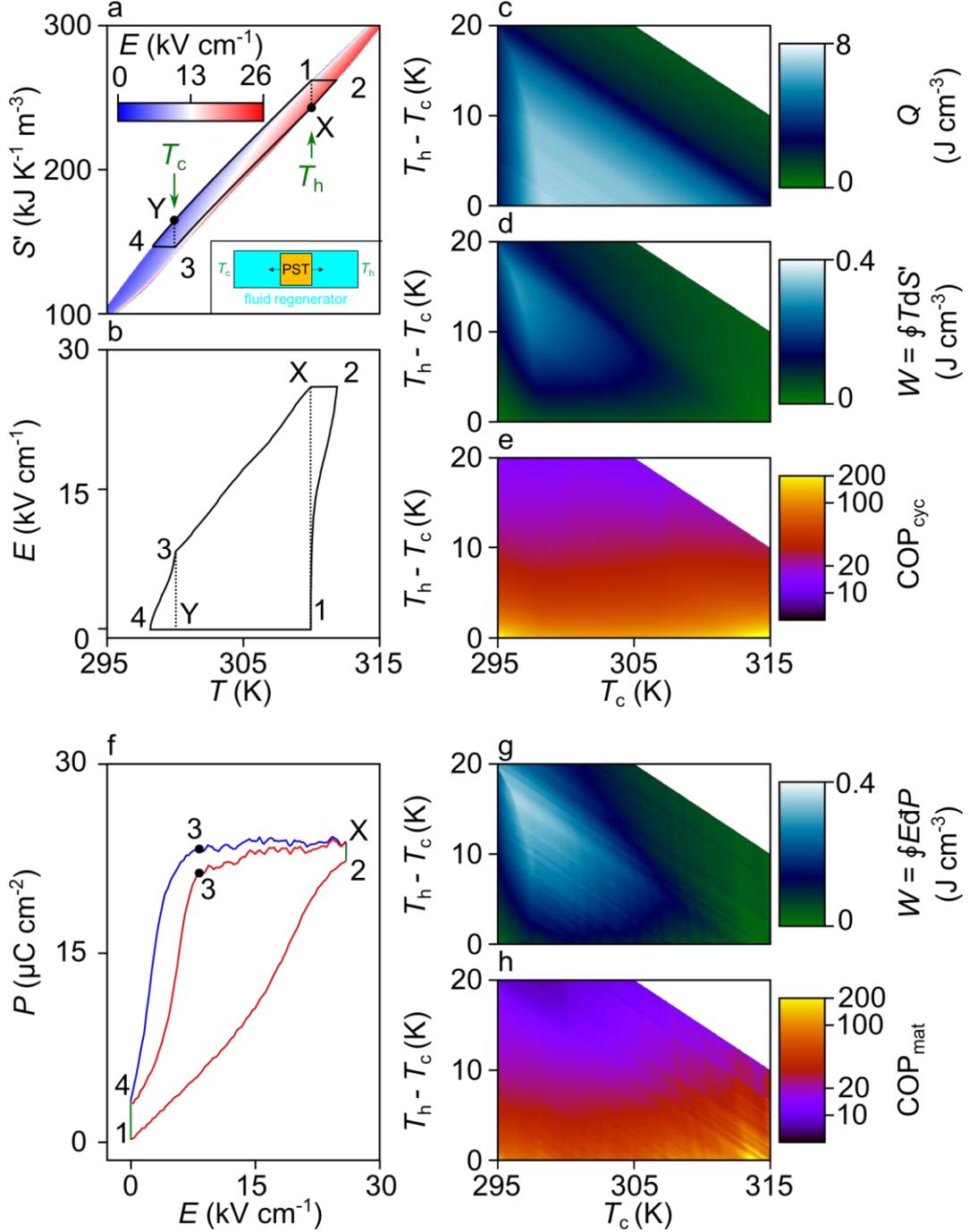

**Figure 4. Electrocaloric cooling cycles with true regeneration for PST above $T_C$.** (a) $E(S',T)$ for $T > T_C$ and $E > 0$, obtained by permuting the variables in $S'(T,E)$ (Fig. 3a). The Brayton-like balanced cooling cycle 1→2→X→3→4→Y→1 with $E_{max} = 26$ kV cm$^{-1}$ assumes the use of an ideal regenerator (inset). Black dotted lines show load temperature $T_c$ and sink temperature $T_h$. (b) The cycle in (a) on $(T,E)$ axes. (c,d,e) On varying $T_c$ and $T_h - T_c$ in our cycle, we plot (c) the heat $Q = \int_4^Y T(S')\,dS'$ absorbed from the load at zero field, (d) cycle work $W = \oint T dS'$, and hence (e) $COP_{cyc} = Q/\oint T dS'$. (f) The cycle in (a) on $(E,P)$ axes, with experimentally obtained data in red, isofields in green, and Y omitted for clarity. Blue data were obtained from the field-removal branches of $|P(S',E)|$ (Fig. 2e), resulting in an expanded cycle to account for field hysteresis. (g,h) On varying $T_c$ and $T_h - T_c$ for cycles thus expanded, we plot (g) work $W = \oint E dP$ and (h) $COP_{mat} = Q/\oint E dP$. Supplementary Fig. 7 repeats (a,b,f) with different values of $T_c$ and $T_h$. Supplementary Fig. 8 repeats (e,h) after dividing by the Carnot limit to obtain efficiency.

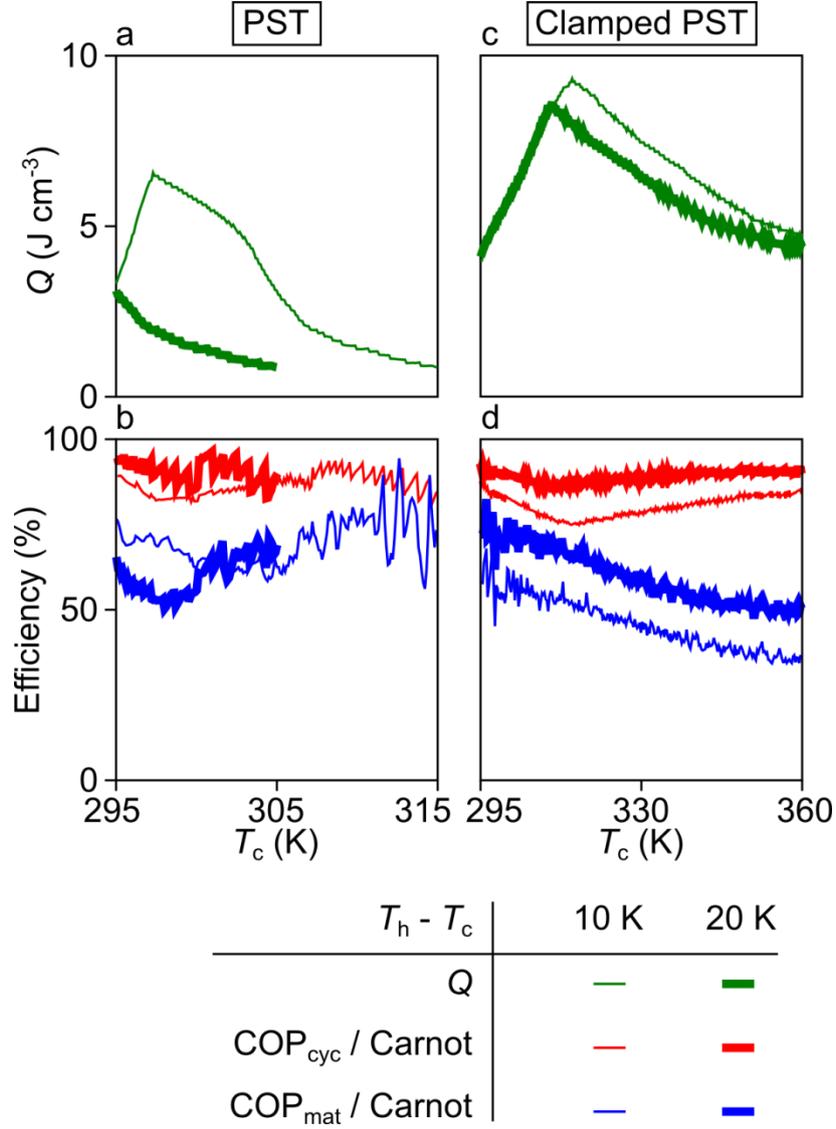

**Figure 5. Heat and efficiency for PST and clamped PST.** (a-d) For $T_h - T_c = 10$ K (thin traces) and 20 K (thick traces), we show (a) heat $Q(T_c)$ (green), and (b) cycle efficiencies $COP_{cyc}(T_c)$/Carnot (red) and $COP_{mat}(T_c)$/Carnot (blue), for (a,b) PST with $E_{max} = 26$ kV cm$^{-1}$ and (c,d) clamped PST with $E_{max} = \pm 160$ kV cm$^{-1}$. Carnot = $T_c/(T_h - T_c)$. Data for PST from Fig. 4c,e,h. Data for clamped PST from Supplementary Figs 15c,e,h. Supplementary Fig. 9 repeats all panels with data in (b,d) multplied by Carnot to obtain $COP_{cyc}(T_c)$ and $COP_{mat}(T_c)$.

# Supplementary Information

for

# Electrocaloric cooling cycles in lead scandium tantalate with true regeneration via field variation

S. Crossley[1], B. Nair[1], R. W. Whatmore[2], X. Moya[1] and N. D. Mathur[1]

[1]Materials Science, University of Cambridge, Cambridge, CB3 0FS, United Kingdom

[2]Department of Materials, Royal School of Mines, South Kensington Campus, Imperial College London, London SW7 2AZ, United Kingdom

## Supplementary Notes

1. Dielectric measurements of PST at selected frequencies
2. All $Q(V)$ plots for PST
3. Maxwell relations
4. Direct EC measurements of PST
5. Cross-sections through Fig. 2 panels
6. Cross-sections through Fig. 3 panels
7. Comparison of indirect/direct values of $|\Delta T|$ for PST on field application/removal
8. Electrocaloric cooling cycles for PST with different values of $T_c$ and $T_h$
9. Repeat of Fig. 4e,h with efficiency instead of COP
10. Repeat of Fig. 5 with COP instead of efficiency
11. Electrical polarization measurements of clamped PST
12. Indirect EC measurements of clamped PST
13. Electrocaloric cooling cycles with true regeneration for clamped PST above $T_C$
14. Electrocaloric cooling cycles for clamped PST with different values of $T_c$ and $T_h$
15. Repeat of Fig. S15e,h with COP instead of efficiency
16. Comparison of EC properties for PST and clamped PST

**Note 1. Dielectric measurements of PST at selected frequencies**

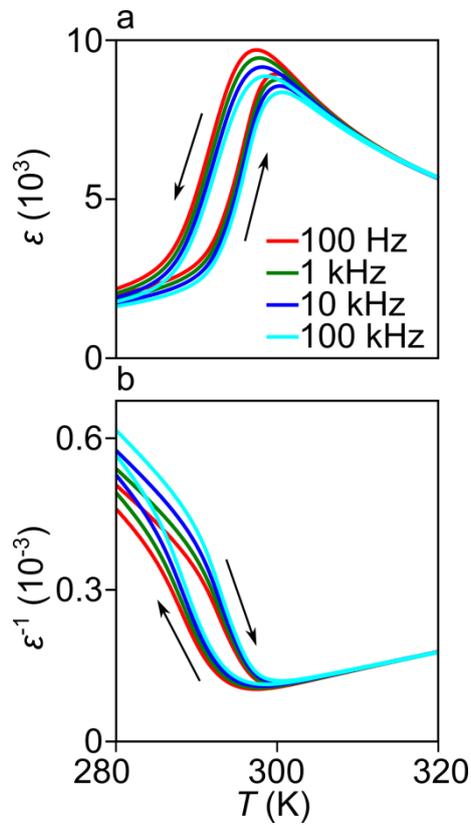

**Figure S1. Dielectric data for PST.** We plot (a) relative dielectric constant $\varepsilon$ and (b) its inverse against temperature $T$, at selected frequencies. Black arrows indicate heating and cooling at ±1 K min$^{-1}$. A small ~1 K dispersion of the dielectric peak is apparent. Data for Sample A.

## Note 2. All *Q*(*V*) plots for PST

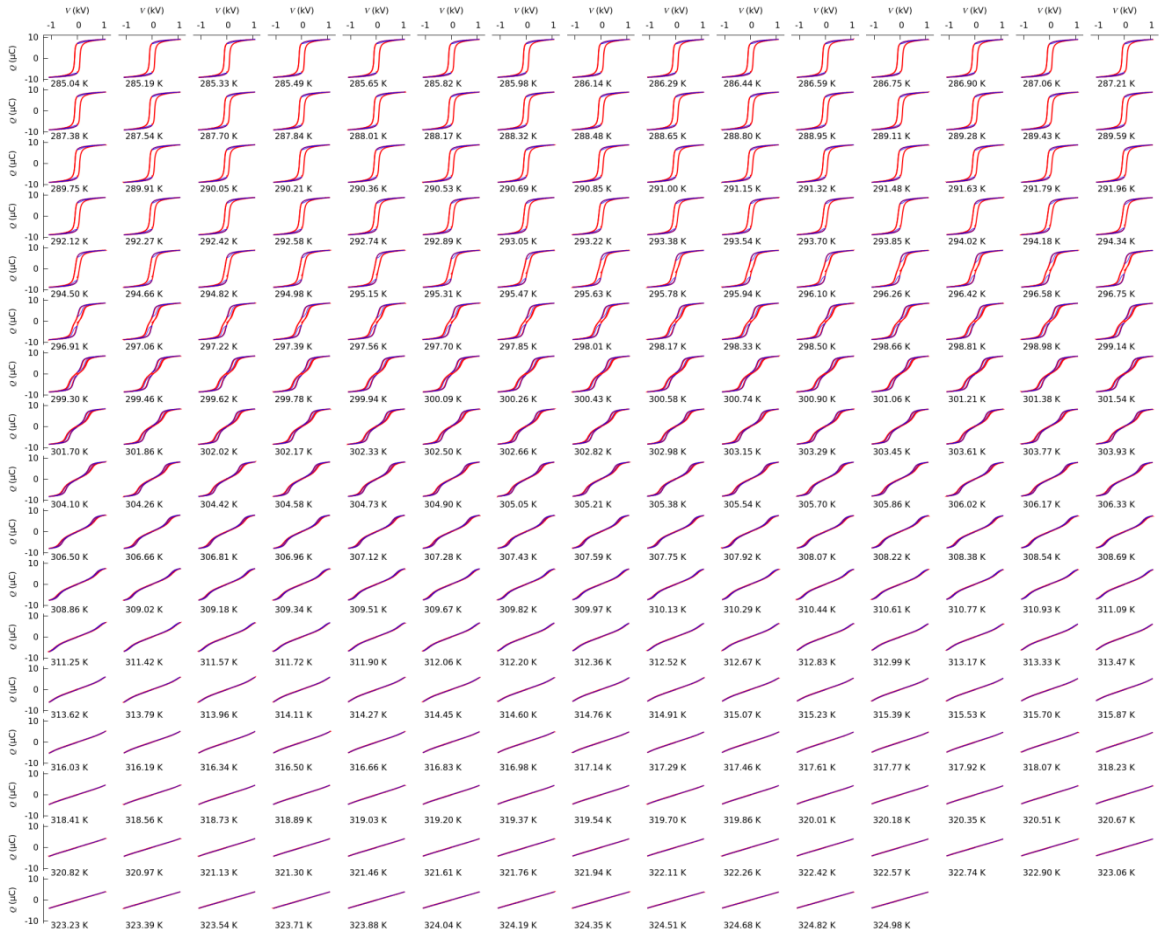

**Figure S2. All *Q*(*V*) plots for PST.** We show at a glance all 252 bipolar (red) and unipolar (blue) plots of charge *Q* versus voltage *V*, which were performed under adiabatic conditions at the start temperatures indicated. Data for Sample A.

# Note 3. Maxwell relations

The Maxwell relation $\left(\frac{\partial S}{\partial E}\right)_T = \left(\frac{\partial P}{\partial T}\right)_E$ derived from Gibbs free energy $G$ is commonly employed in order to evaluate EC effects from electrical polarization data obtained under isothermal conditions. However, electrical polarization data in our paper were obtained under adiabatic conditions, and so instead we employ the Maxwell relation $\left(\frac{\partial T}{\partial E}\right)_S = -\left(\frac{\partial P}{\partial S}\right)_E$ derived from enthalpy $H$.

**Maxwell relation from Gibbs free energy**

For changes in Gibbs free energy $G$:

$$dG = -SdT - DdE$$

where $S$ is entropy, $T$ is temperature, $D$ is dielectric displacement and $E$ is electric field, we have:

$$-S = \left(\frac{\partial G}{\partial T}\right)_E \qquad -D = \left(\frac{\partial G}{\partial E}\right)_T$$

such that:

$$-\frac{\partial^2 G}{\partial T \partial E} = \left(\frac{\partial S}{\partial E}\right)_T = \left(\frac{\partial D}{\partial T}\right)_E$$

or equally:

$$\boxed{\left(\frac{\partial S}{\partial E}\right)_T = \left(\frac{\partial P}{\partial T}\right)_E}$$

given that $\varepsilon_0 E = D - P$ implies $\left(\frac{\partial D}{\partial T}\right)_E = \left(\frac{\partial P}{\partial T}\right)_E$.

**Maxwell relation from enthalpy**

For changes in enthalpy $H$:

$$dH = TdS - DdE$$

we have:

$$T = \left(\frac{\partial H}{\partial S}\right)_E \qquad -D = \left(\frac{\partial H}{\partial E}\right)_S$$

such that:

$$\frac{\partial^2 H}{\partial S \partial E} = \left(\frac{\partial T}{\partial E}\right)_S = -\left(\frac{\partial D}{\partial S}\right)_E$$

or equally:

$$\boxed{\left(\frac{\partial T}{\partial E}\right)_S = -\left(\frac{\partial P}{\partial S}\right)_E}$$

given that $\varepsilon_0 E = D - P$ implies $\left(\frac{\partial D}{\partial S}\right)_E = \left(\frac{\partial P}{\partial S}\right)_E$.

## Note 4. Direct EC measurements of PST

Non-adiabatic thermocouple measurements of EC temperature change, at a series of decreasing starting temperatures, are presented in the adiabatic limit (Fig. S3) after calibration with the highly adiabatic scanning thermal microscopy (SThM) measurement of EC temperature change shown in Fig. S4a (which was approximately consistent with the infra-red thermometry shown in Fig. S4b).

Figs S3 & S4 appear on the next two sides.

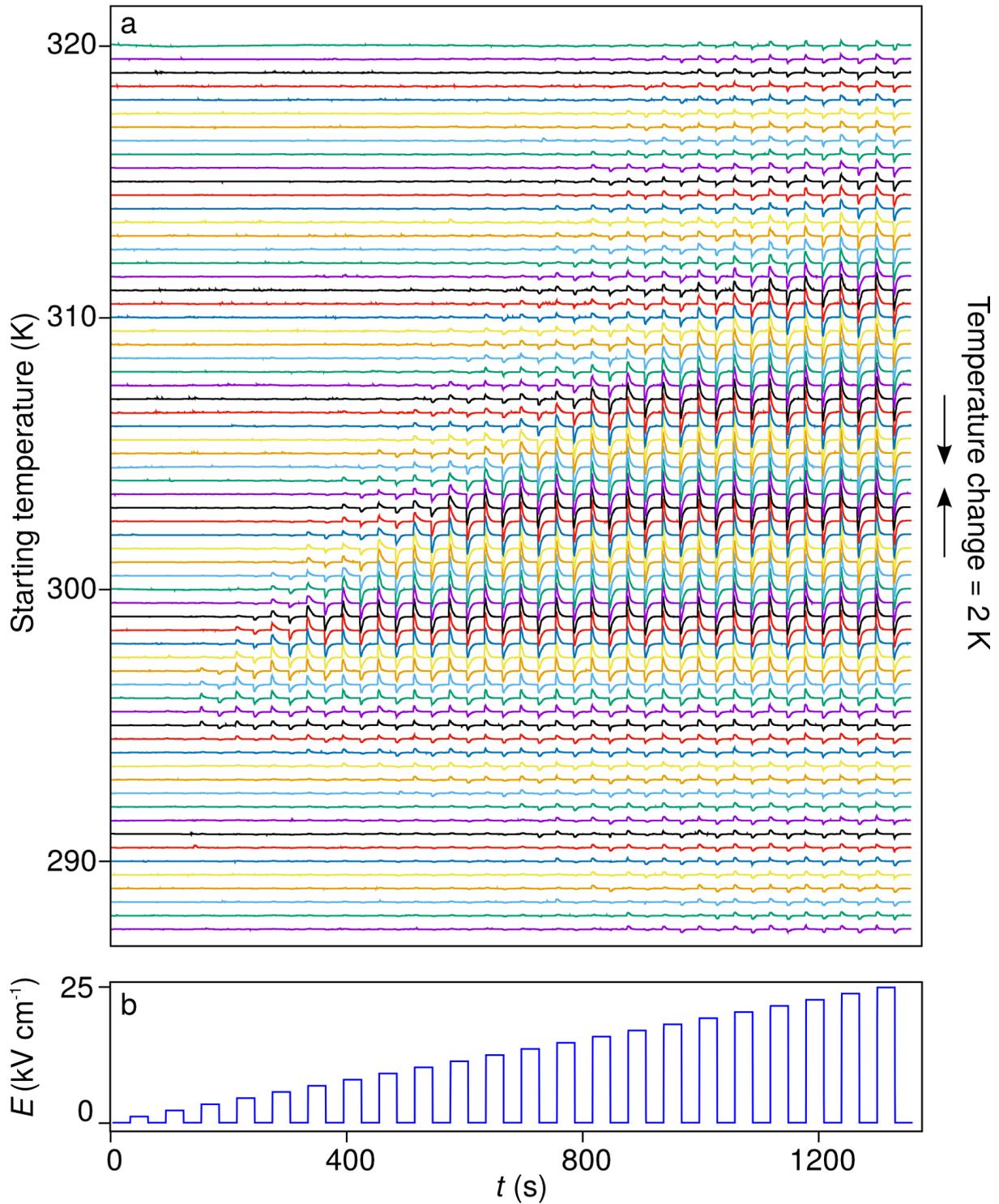

**Figure S3. Thermocouple measurements of temperature change in PST starting at various temperatures.** (a) At a series of decreasing starting temperatures (left axis), we show changes of temperature (right axis) versus time $t$ due to (b) changes of electric field $E(t)$. These changes of temperature were evaluated from non-adiabatic changes in measured thermoelectric voltage, using a calibration factor (62.97 K mV$^{-1}$) that we established via the highly adiabatic SThM data of Fig. S4a (See Methods in the main paper). The jumps shown here therefore represent adiabatic changes of temperature $\Delta T$, and were used to construct maps in the main paper showing $|\Delta T(S',E)|$ (Fig. 2k,o), $T(S',E)$ (Fig. 2l,p), and also $S'(T,E)$ (Fig. 3i,m) where temperature change is apparent from contours. Data for Sample C.

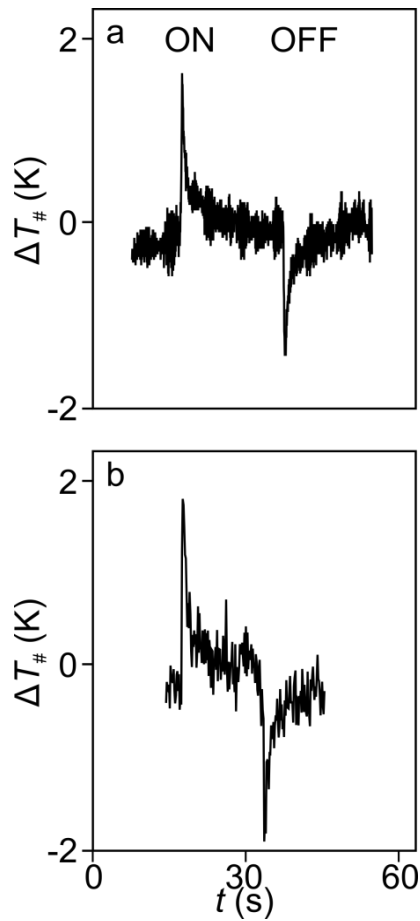

**Figure S4. SThM and infra-red measurements of temperature change in PST starting at room temperature.** Under highly adiabatic conditions, (a) SThM and (b) infra-red thermometry reveal temperature change $\Delta T_\#$ versus time $t$ when turning on and off an electric field of 15 kV cm$^{-1}$. The sharp jumps are assumed to represent adiabatic changes of temperature $\Delta T$. Data at ~297 K for (a) Sample B and (b) Sample A.

## Note 5. Cross-sections through Fig. 2 panels

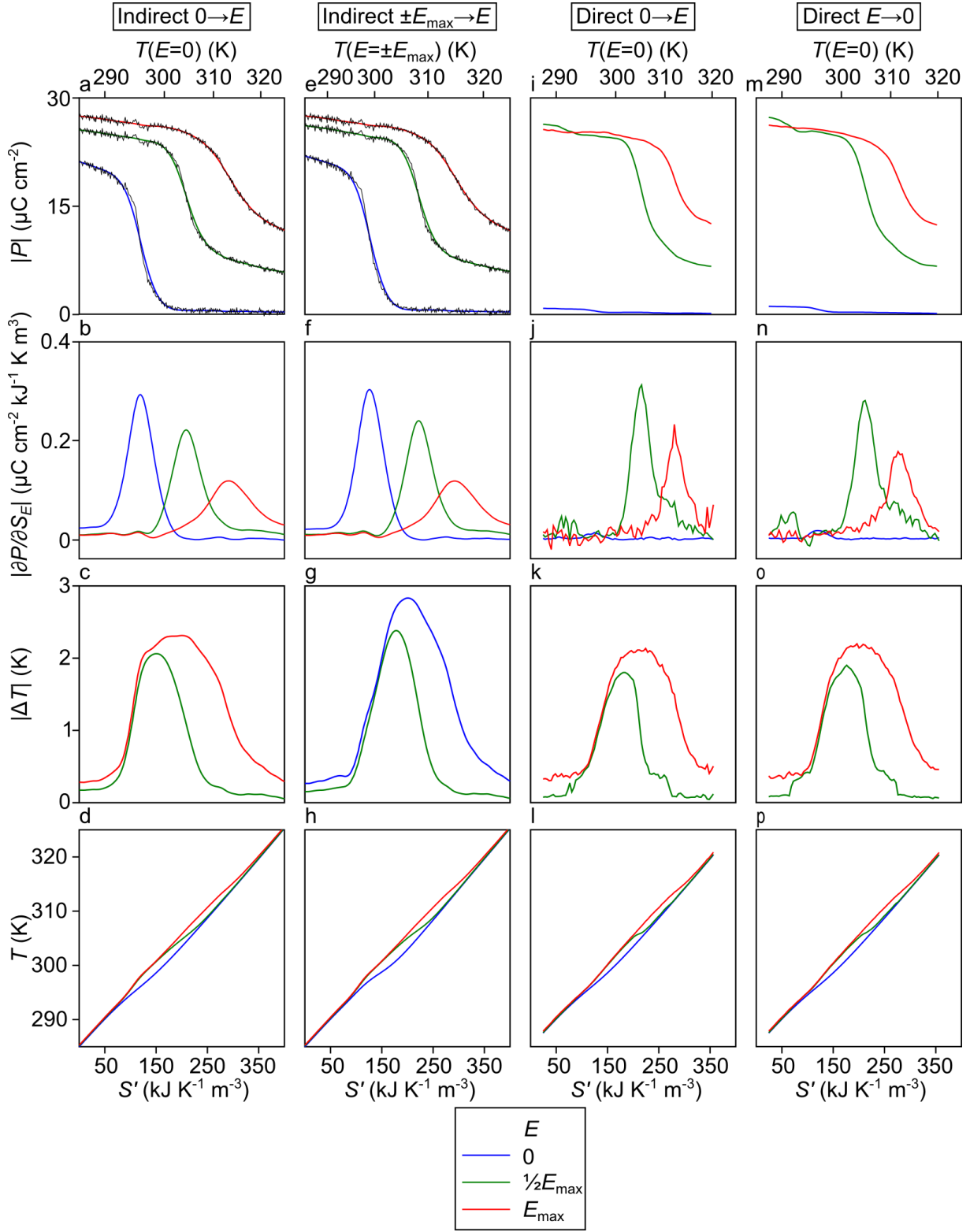

**Figure S5. Cross-sections through Fig. 2 panels.** Cross-sections are presented for $E = 0$ (blue), $\tfrac{1}{2}E_{max}$ (green) and $E_{max}$ (red), where (a-h) $E_{max} = 26$ kV cm$^{-1}$ and (i-p) $E_{max} = 24.5$ kV cm$^{-1}$. Thin black lines in (a,e) denote data prior to spline fitting. Boxed text at top describes method of measurement.

## Note 6. Cross-sections through Fig. 3 panels

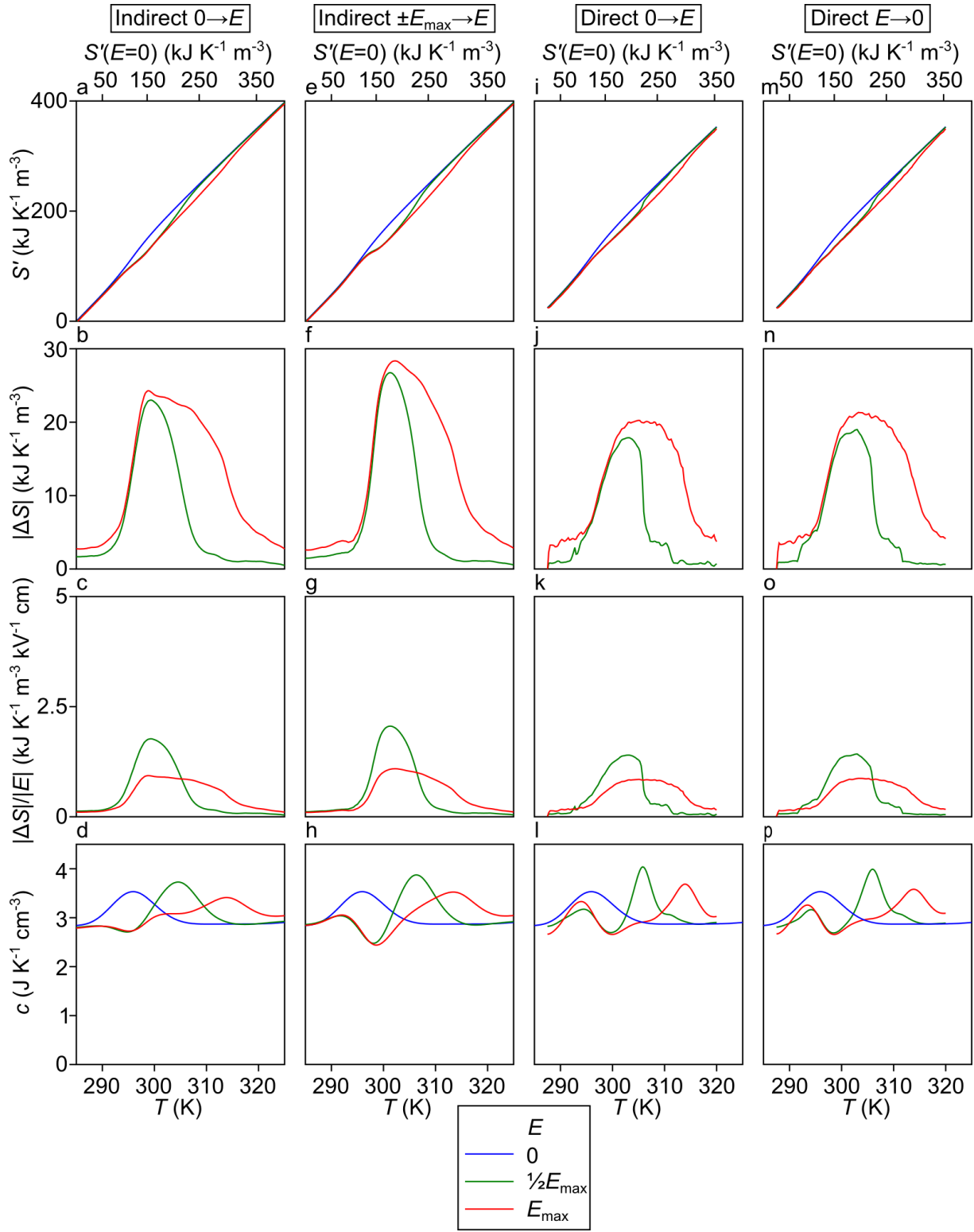

**Figure S6. Cross-sections through Fig. 3 panels.** Cross-sections are presented for $E = ¼E_{max}$ (black), $½E_{max}$ (green) and $E_{max}$ (red), where (a-h) $E_{max} = 26$ kV cm$^{-1}$ and (i-p) $E_{max} = 24.5$ kV cm$^{-1}$. Boxed text at top describes method of measurement.

**Note 7. Comparison of indirect/direct values of |Δ*T*| for PST on field application/removal**

The reader may compare our indirect/direct EC maps for field application/removal (Figs 2,3 in the main paper) by viewing the constant-field cross-sections in Figs S5-6. The small discrepancies that we observe do not arise from sample-to-sample variations, as all samples came from a single wafer whose fabrication process implies intrinsically uniform composition and cation ordering.

**Comparison of indirect EC measurements of |Δ*T*| for field application and field removal**

The peak indirect value of |Δ*T*| ~ 2.8 K obtained from unipolar field-removal branches (Fig. S5g) slightly exceeds its |Δ*T*| ~ 2.3 K counterpart obtained from unipolar field-application branches (Fig. S5c) because the single-valued assumption on $P(S',E)$ is slightly compromised by the small field hysteresis that arises due to pinning when driving the transition (Fig. 1d in the main paper). Importantly, the heating produced by the small field hysteresis is not significant, as the area enclosed by a representative unipolar loop (305 K data in $E > 0$, Fig. 1d in the main paper) implies a temperature increase of just ~12 mK.

**Comparison of indirect and direct EC measurements of |Δ*T*|**

The peak direct value of |Δ*T*| ~ 2.2 K for both EC heating (Fig. S5k) and cooling (Fig. S5o) matches well the peak indirect value of |Δ*T*| ~ 2.3 K for unipolar field-application (Fig. S5g). As explained in Methods, the resolution of these direct EC measurements is compromised by the finite heat capacities of the thermocouple and top electrode, with two consequences. First, we do not resolve the small field hysteresis of the transition (Fig. 1d in the main paper). Second, we do not resolve small EC effects, such that the transition appears to be sharpened, thus separating the 'teeth' in Fig. 2k,o of the main paper.

**Note 8. Electrocaloric cooling cycles for PST with different values of $T_c$ and $T_h$**

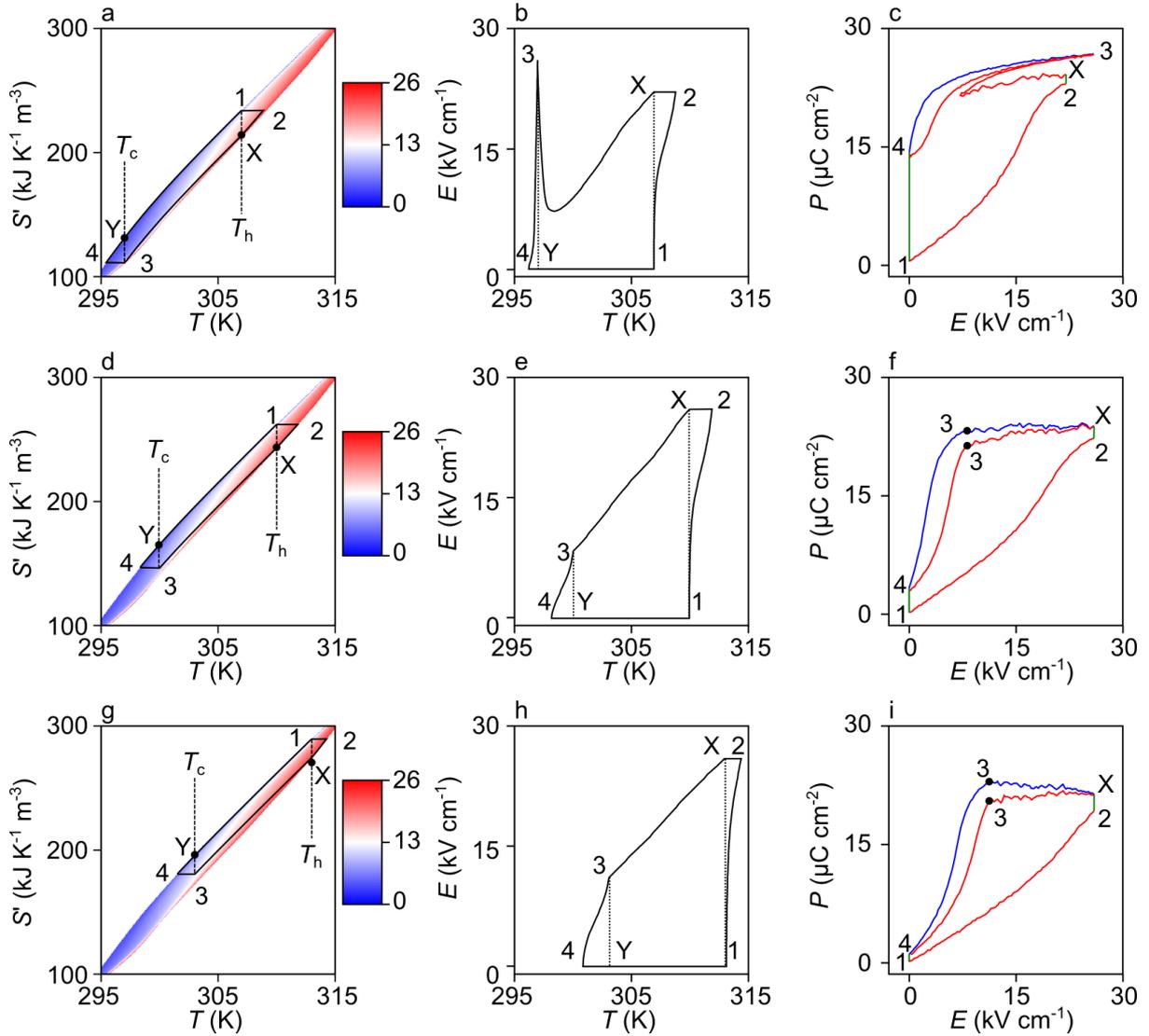

**Figure S7. Representative Brayton-like cycles with true regeneration for PST above $T_C$.** The panels of Fig. 4a,b,f in the main paper are redrawn for cycles with (a-c) low, (d-f) intermediate and (g-i) high values of $T_c$ and $T_h$. Each cycle reaches $E_{max} = 26$ kV cm$^{-1}$ during 2-X-3. Non-monotonicity in (b,c) evidences a small mismatch between absolute temperatures in the underlying electrical and thermal data.

**Note 9. Repeat of Fig. 4e,h with efficiency instead of COP**

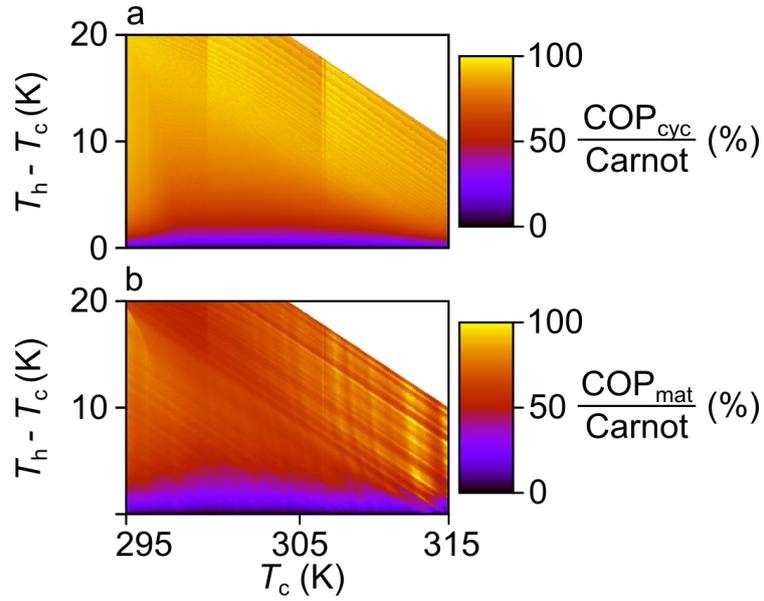

**Figure S8. COP$_{cyc}$ and COP$_{mat}$ as fractions of the Carnot limit.** (a) COP$_{cyc}$ = $Q/\oint T dS'$ (Fig. 4e in the main paper) is presented here after dividing through by the Carnot limit $T_c/(T_h - T_c)$. (b) COP$_{mat}$ = $Q/\oint E\text{đ}P$ (Fig. 4h in the main paper) is presented here after dividing through by the Carnot limit $T_c/(T_h - T_c)$. Cross-sections through each panel at $T_h - T_c$ = 10 K and 20 K appear in Fig. 5b in the main paper.

**Note 10. Repeat of Fig. 5 with COP instead of efficiency**

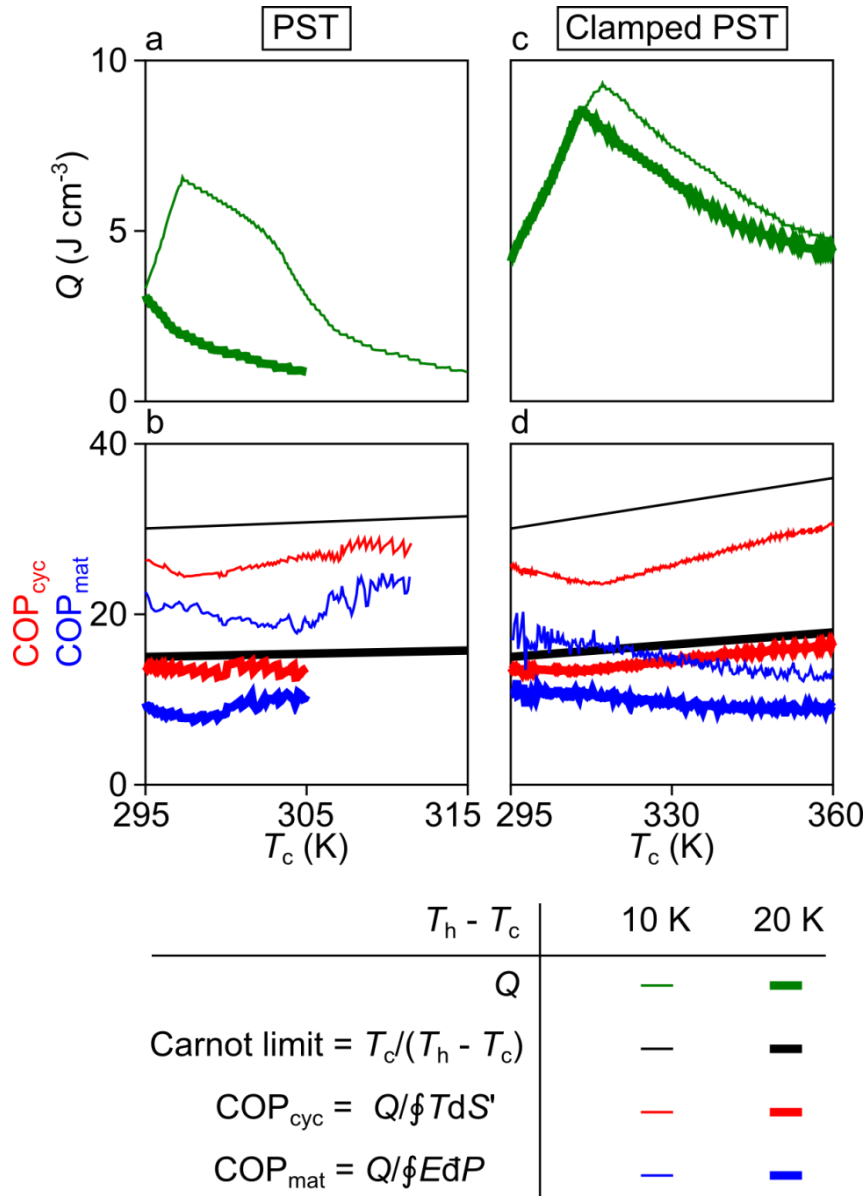

**Figure S9. Heat and coefficients of performance for PST and clamped PST.** (a,b) For $T_h - T_c = 10$ K (thin traces) and 20 K (thick traces), we show (a) heat $Q(T_c)$ (green), and (b) $COP_{cyc}(T_c)$ (red), $COP_{mat}(T_c)$ (blue) and the Carnot limit $T_c/(T_h - T_c)$ (black), for PST with $E_{max} = 26$ kV cm$^{-1}$ and (c,d) clamped PST with $E_{max} = \pm 160$ kV cm$^{-1}$. Data for PST from Fig. 4c,e,h. Data for clamped PST from Supplementary Figs 15c,e,h. Fig. 5 in the main paper repeats all panels with data in (b,d) divided by the Carnot limit to obtain cycle efficiencies $COP_{cyc}(T_c)$/Carnot and $COP_{mat}(T_c)$/Carnot.

## Note 11. Electrical polarization measurements of clamped PST

Clamped PST with a small top electrode was measured in a similar manner to PST, using ~0.26 K intervals, and field steps of ~1.9 kV cm$^{-1}$ up to a larger value of $E_{max}$ = 160 kV cm$^{-1}$. We show four representative bipolar (Fig. S10a) and unipolar (Fig. S10b) plots that form part of the full dataset obtained at 438 measurement set temperatures (Fig. S11). The diffuse nature of the electrically and thermally driven transitions is attributed to mechanical clamping. This clamping arises primarily due to the relatively large volume of unaddressed material, and perhaps partly due to enhanced substrate clamping on account of reduced sample thickness.

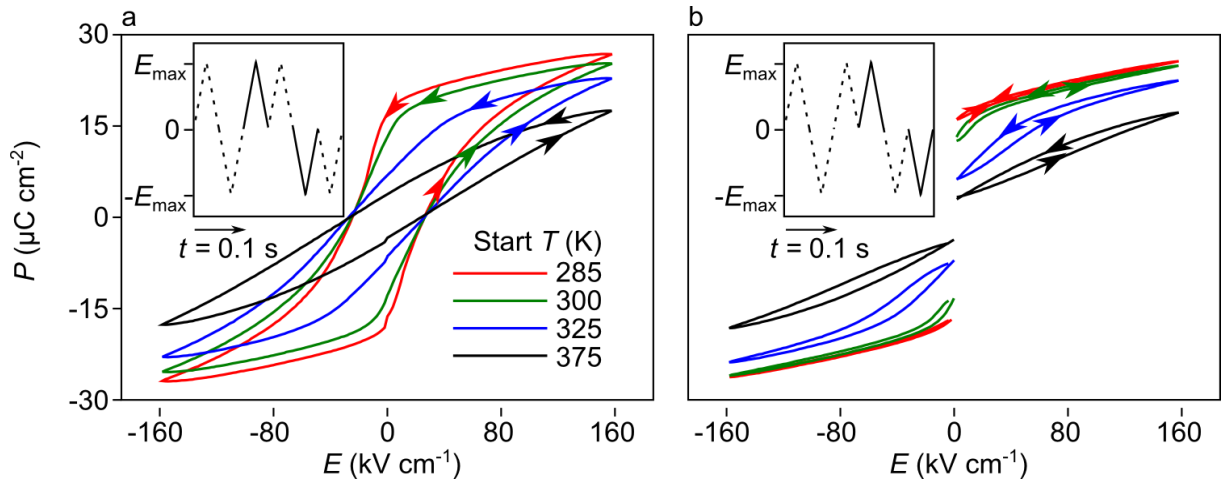

**Figure S10. Electrical polarization in clamped PST.** Adiabatic polarization $P(E)$ measured with $E_{max}$ = 160 kV cm$^{-1}$ at selected starting temperatures on heating. Data for (a) bipolar and (b) unipolar plots were acquired during the times denoted by solid lines on the insets, which show driving field $E$ versus time $t$. Raw data at all 438 measurement temperatures appear in Fig. S11. Data for Sample D.

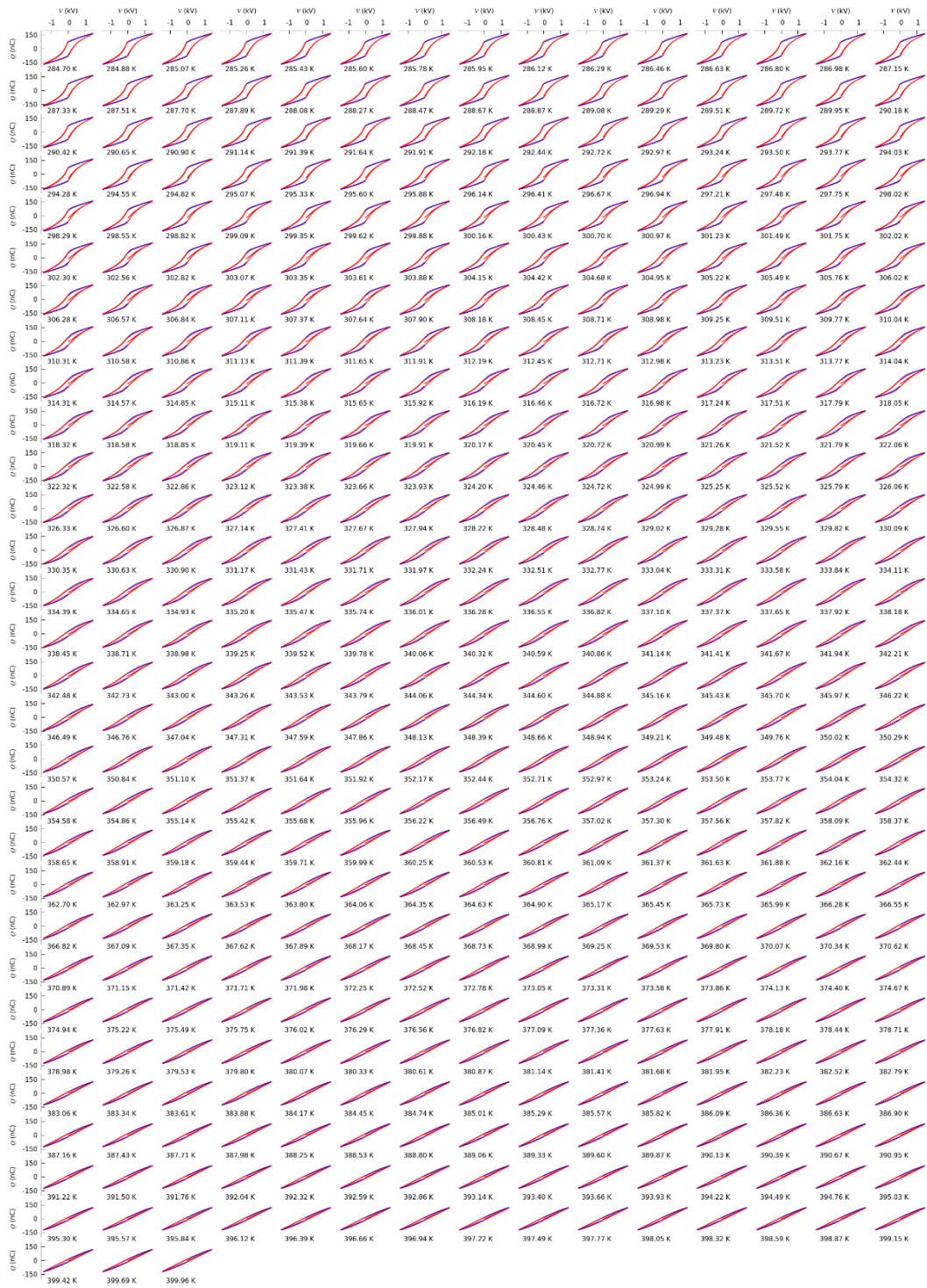

**Figure S11. All *Q*(*V*) plots for clamped PST.** We show at a glance all 438 bipolar (red) and unipolar (blue) plots of charge *Q* versus voltage *V*, which were performed under adiabatic conditions at the start temperatures indicated. Data for Sample D, prior to the leakage correction described in this Supplementary Note.

As described below, clamped PST (Sample D) required initial electrical conditioning, and displayed a small electrical leakage that we corrected (along with a small parasitic capacitance) by comparing bipolar and unipolar electrical polarization data. Both the conditioning and leakage issues are understood to be an artefact of hand thinning. Given that MLCs with even thinner layers show no high-field ohmic leakage[S1], the small electrical leakage displayed by Sample D is irrelevant for applications.

**Electrical conditioning of clamped PST**

For clamped PST, high voltages initially led to high conductivity, but insulating behaviour could be recovered by heating above 320 K, possibly due to a neutralising flow of the Apiezon 'N' grease that covers the top electrode into pinhole defects (the grease flows more easily above room temperature). After repeating this process a number of times, stable insulating behaviour was observed for applied fields below $E_{max}$ ~ 160 kV cm$^{-1}$.

**Correction for ohmic leakage and parasitic capacitance of clamped PST**

**Step 1) Correction for small ohmic leakage by comparing bipolar and unipolar data.** Electrical polarization data are deduced from measurements of charge $Q$ versus voltage $V$, where bipolar and unipolar conditions yield slightly different values of $Q$ near and at the maximum positive applied voltage $+V_{max}$ (Fig. S12a-e) and the maximum negative applied voltage $-V_{max}$ (not shown). This discrepancy may be understood by assuming that a parallel resistance $R$ results in a leakage current that is directly proportional to applied voltage $V(t)$, such that the net charge that flowed through the resistance is given by $Q_{leak} = \int V(t)/R \, dt$ (this integration is performed automatically by the Radiant tester, $t$ is time). In a bipolar (unipolar) cycle that starts and finishes at $+V_{max}$, the value of $Q_{leak}$ is zero (finite). We therefore identify $Q_{leak}$ via the discrepancy between our bipolar and unipolar $Q(V)$ plots at $+V_{max}$. At each of our 438 measurement temperatures, values of $Q_{leak}$ are smaller for $+V_{max}$ than their counterparts identified at $-V_{max}$ (Fig. S12f), so we have used the smaller values obtained at $+V_{max}$ to avoid overestimating the correction. Each value of $Q_{leak}$ yields a value of $R$ that permits us to plot the net charge that has flowed through this resistance during the corresponding bipolar cycle, with $Q = \pm Q_{leak}$ at $V = 0$ (pink, Fig. S12a-e). The net charge that has flowed through the resistance during each unipolar cycle may be plotted similarly (not shown). At 305 K, where $+Q_{leak}$ is largest, we find $R$ = 14 GΩ, corresponding to a resistivity of 4.9×10$^9$ Ω cm.

**Step 2) Correction for small ohmic leakage and parasitic capacitance.** After thus correcting all bipolar $Q(V)$ data for leakage, we identified a temperature-independent parasitic capacitance $Q/V = C = 38$ pF from the finite gradient of upper branches on the approach to $+V_{max}$. This parasitic capacitance is thought to arise from the wiring in the cryostat. The combined contribution (black, Fig. S12a-d) of leakage current and parasitic capacitance were used to correct all bipolar and unipolar electrical polarization data for clamped PST that appear elsewhere in this paper (green data in Fig. S12a-d show corrected bipolar $Q(V)$ plots).

**No correction required for PST**

Leakage and parasitic capacitance have negligible effect on our electrical polarization measurements of PST, which are uncorrected everywhere except for Fig. S12g, where the parasitic capacitance ($C = 38$ pF) is used to make a very small correction in order to identify as precisely as possible $P_s(285\ K) = 26.1\ \mu C\ cm^{-2}$, for use below.

**Effective area of top electrode for clamped PST**

The nominal area of the small top Pt electrode (~0.68 mm$^2$) was not used to calculate polarization $P$ from charge $Q$, in view of physical processes that may have taken place during the aforementioned electrical conditioning. Instead, we established an effective top electrode area of 0.42 mm$^2$ by assuming that our clamped PST with saturation charge $Q_s(285\ K) = 110$ nC (Fig. S12h, data copied from Fig. S12a) possesses the same saturation polarization $P_s(285\ K) = 26.1\ \mu C\ cm^{-2}$ as PST.

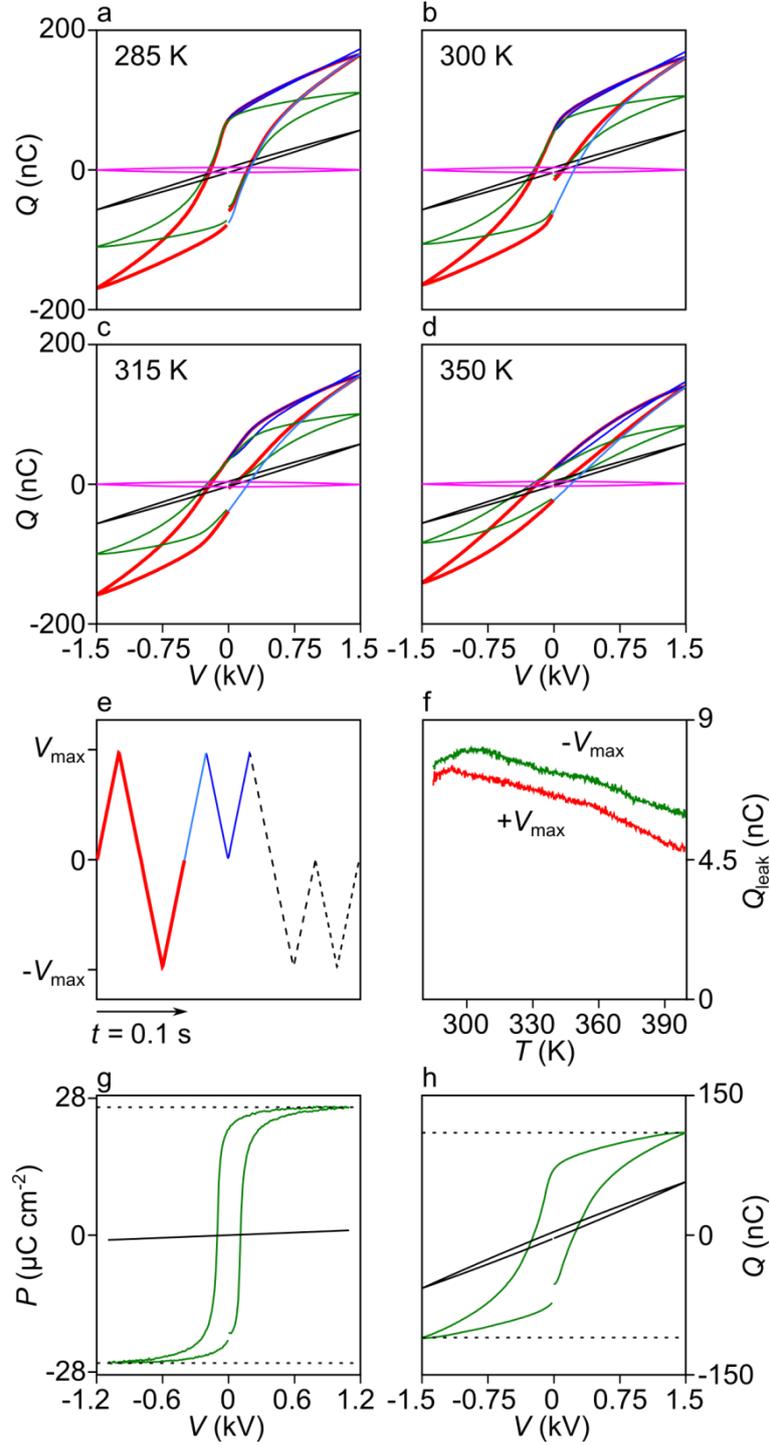

**Figure S12. Ohmic leakage and parasitic capacitance for clamped PST.** At selected temperatures, we show (a-d) charge $Q$ versus voltage $V$, where a complete bipolar cycle ($0 \to V_{max} \to -V_{max} \to 0$, red) continues ($0 \to V_{max}$, light blue) prior to a unipolar cycle ($V_{max} \to 0 \to V_{max}$, dark blue). The triangular profile of the applied voltage is shown using the same colour code in (e). In (a-d), the temperature-dependent Ohmic contribution (pink) is combined with the temperature-independent parasitic capacitance of 38 pF (straight line, not shown) to yield a combined contribution (black) that is subtracted to yield corrected bipolar $Q(V)$ plots (green). (f) $Q_{leak}$ versus temperature $T$, as measured at $+V_{max}$ and $-V_{max}$. (g) Bipolar $P(V)$ for PST at 285 K (green), after subtracting $C = 38$ pF (black) to obtain $|P_s| = 26.1$ µC cm$^{-2}$ (dotted lines). (h) Bipolar $Q(V)$ for clamped PST at 285 K (green), after subtracting for both (black) parasitic capacitance and Ohmic conduction to obtain $|Q_s| = 110$ nC (dotted lines). Data in (h) are copied from (a).

**Note 12. Indirect EC measurements of clamped PST**

Indirect EC measurements of clamped PST yield entropy-field maps of polarization and temperature (Fig. S13) and temperature-field maps of entropy (Fig. S14), cf. Figs 2&3 in the main paper for PST.

The phase transition for clamped PST is observed at $T_C \sim 295$ K as for PST, but the clamping broadens it to higher temperatures, and the reduced sample volume permits the application of a higher field (Fig. 2 axes for PST in the main paper are denoted by the red dotted line in Fig. S13c).

The largest EC effects in clamped PST are $|\Delta T| \sim 3.3$ K and $|\Delta S| \sim 29.8$ kJ K$^{-1}$ m$^{-3}$ for the application or removal of $E_{max} = 160$ kV cm$^{-1}$.

Figs S13 & S14 appear on the next two sides.

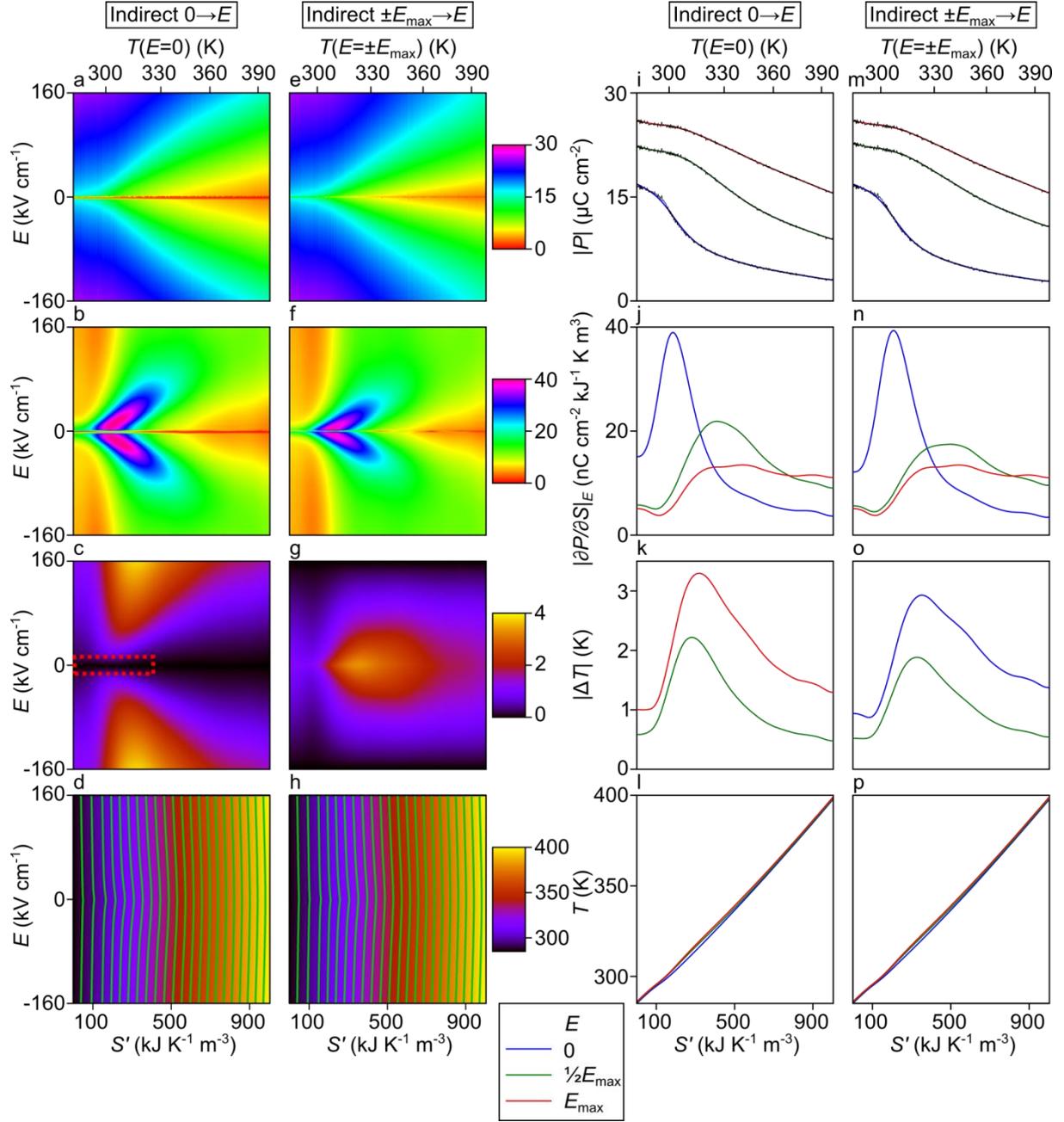

**Figure S13. Polarization and temperature on axes of entropy and field for clamped PST, and cross-sections.** Data based on indirect EC measurements for (a-d) $0 \to E$ and (e-h) $\pm E_{max} \to E$. For each sign of field, $|P(S',E)|$ was constructed by plotting 100 isofield cubic spline fits $P(S')$ to (a) field-application branches ($0 \to E$) and (e) field-removal branches ($\pm E_{max} \to E$) of 438 unipolar $P(E)$ plots (Fig. S11) obtained at starting temperatures separated by 0.26 K. The resulting plots of (b,f) $|\partial P/\partial S|_E$ imply nominally reversible adiabatic temperature changes of (c,g) $|\Delta T(S',E)|$ starting at (c) $T(E=0)$ and (g) $T(E=\pm E_{max})$. Hence (d,h) $T(S',E)$, with isothermal contours every ~4.6 K. Red dotted line in (c) denotes the corresponding axes for PST in Fig. 2c in the main paper. Data for Sample D. (i-p) Constant-field cross-sections through (a-h) for $E = 0$ (blue), $½E_{max}$ (green) and $E_{max} = 160$ kV cm$^{-1}$ (red). Data underlying the spline fits in (i,m) appear black. Variation of starting temperature $T(E=0)$ (top abscissa) with entropy $S'$ (bottom abscissa) from $S'(T)$ for unclamped PST (Fig. 1b in the main paper), which is reasonable because background entropy dominates,.

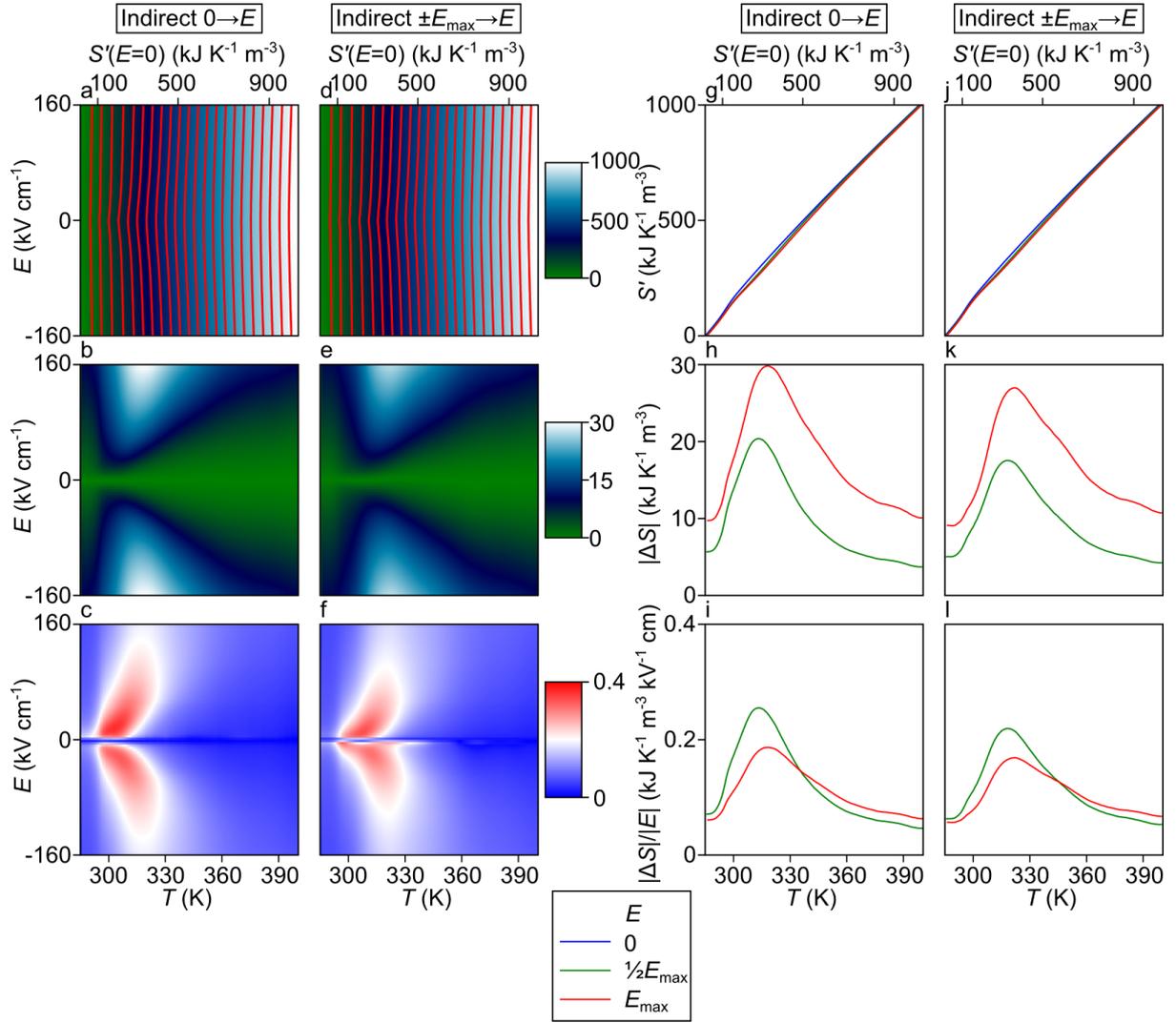

**Figure S14. Entropy on axes of temperature and field for clamped PST, and cross-sections.** (a,d) Entropy $S'(T,E)$ obtained by permuting the variables in $T(S',E)$ (Fig. S13d,h) yields, with adiabatic contours every ~44 kJ K$^{-1}$ m$^{-3}$. Hence (b,e) the nominally reversible isothermal entropy change $|\Delta S(T,E)|$ for $0 \leftrightarrow E$ at temperature $T$. (c,f) EC strength $|\Delta S(T,E)|/|E|$. Data for Sample D. (g-l) Constant-field cross-sections through (a-f) for $E = 0$ (blue), ½$E_{max}$ (green) and $E_{max}$ = 160 kV cm$^{-1}$ (red).

**Note 13. Electrocaloric cooling cycles with true regeneration for clamped PST above $T_C$**

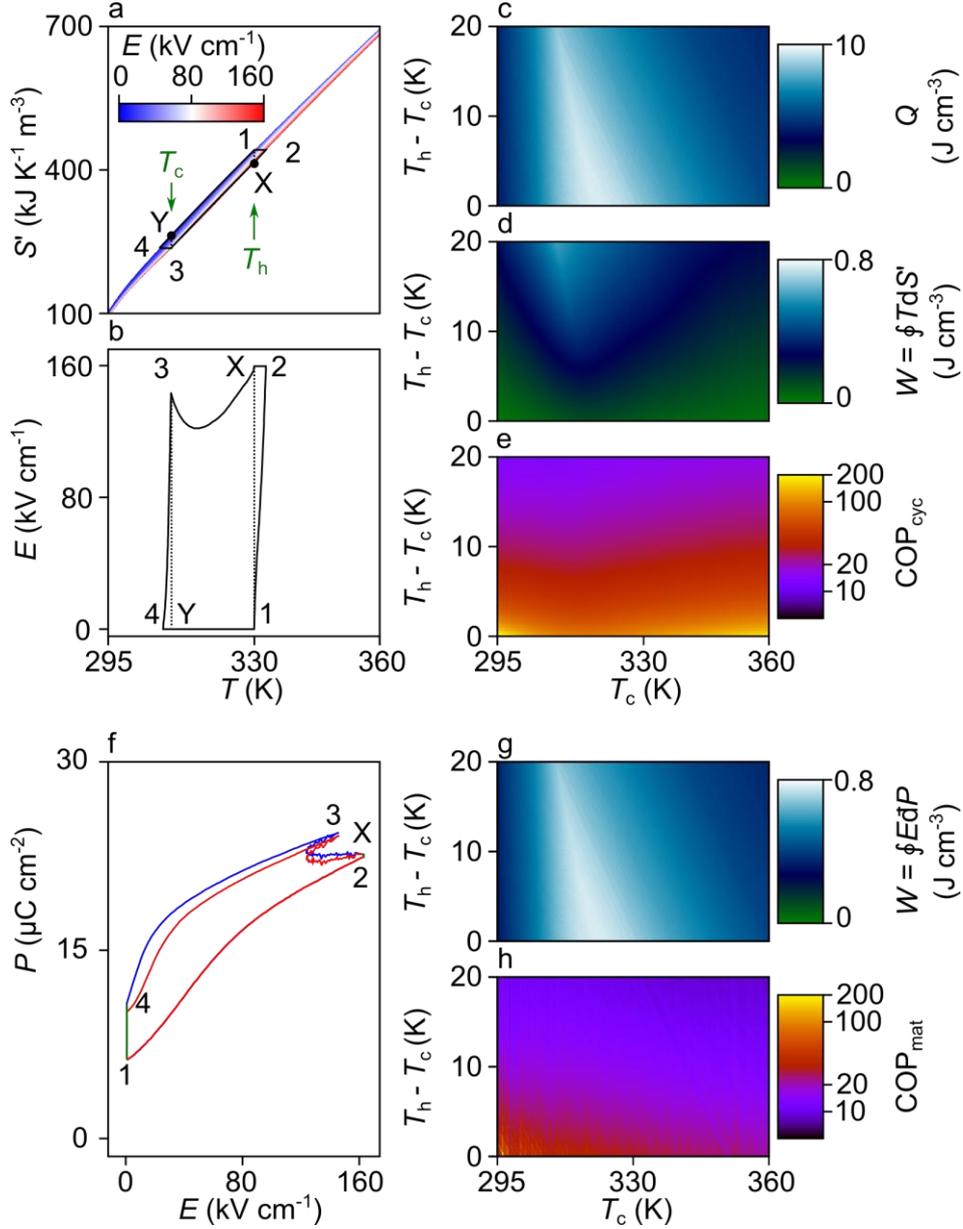

**Figure S15. Electrocaloric cooling cycles with true regeneration for clamped PST above $T_C$.** (a) $E(S',T)$ for $T > T_C$ and $E > 0$, obtained by permuting the variables in $S'(T,E)$ (Fig. S14a). The Brayton-like balanced cooling cycle $1\to2\to X\to3\to4\to Y\to1$ with $E_{max} = 160$ kV cm$^{-1}$ assumes the use of an idealised regenerator. Black dotted lines show load temperature $T_c$ and sink temperature $T_h$. (b) The cycle in (a) on $(T,E)$ axes. (c,d,e) On varying $T_c$ and $T_h - T_c$ in our cycle, we plot (c) the heat $Q = \int_4^Y T(S')\,dS'$ absorbed from the load at zero field, (d) cycle work $W = \oint T dS'$ and hence (e) $COP_{cyc} = Q/\oint T dS'$. (f) The cycle in (a) on $(E,P)$ axes, with experimentally obtained data in red, isofields in green, and Y omitted for clarity. Blue data were obtained from the field-removal branches of $|P(S',E)|$ (Fig. S13e), resulting in an expanded cycle to account for field hysteresis. (g,h) On varying $T_c$ and $T_h - T_c$ for cycles thus expanded, we plot (g) work $W = \oint E dP$ and (h) $COP_{mat} = Q/\oint E dP$. Fig. S16 repeats (a,b,f) with different values of $T_c$ and $T_h$. Fig. S17 repeats (e,h) after dividing by the Carnot limit to obtain efficiency. Non-monotonicity in (b,f) evidences a small mismatch between absolute temperatures in the underlying electrical and thermal data.

**Note 14. Electrocaloric cooling cycles for clamped PST with different values of $T_c$ and $T_h$**

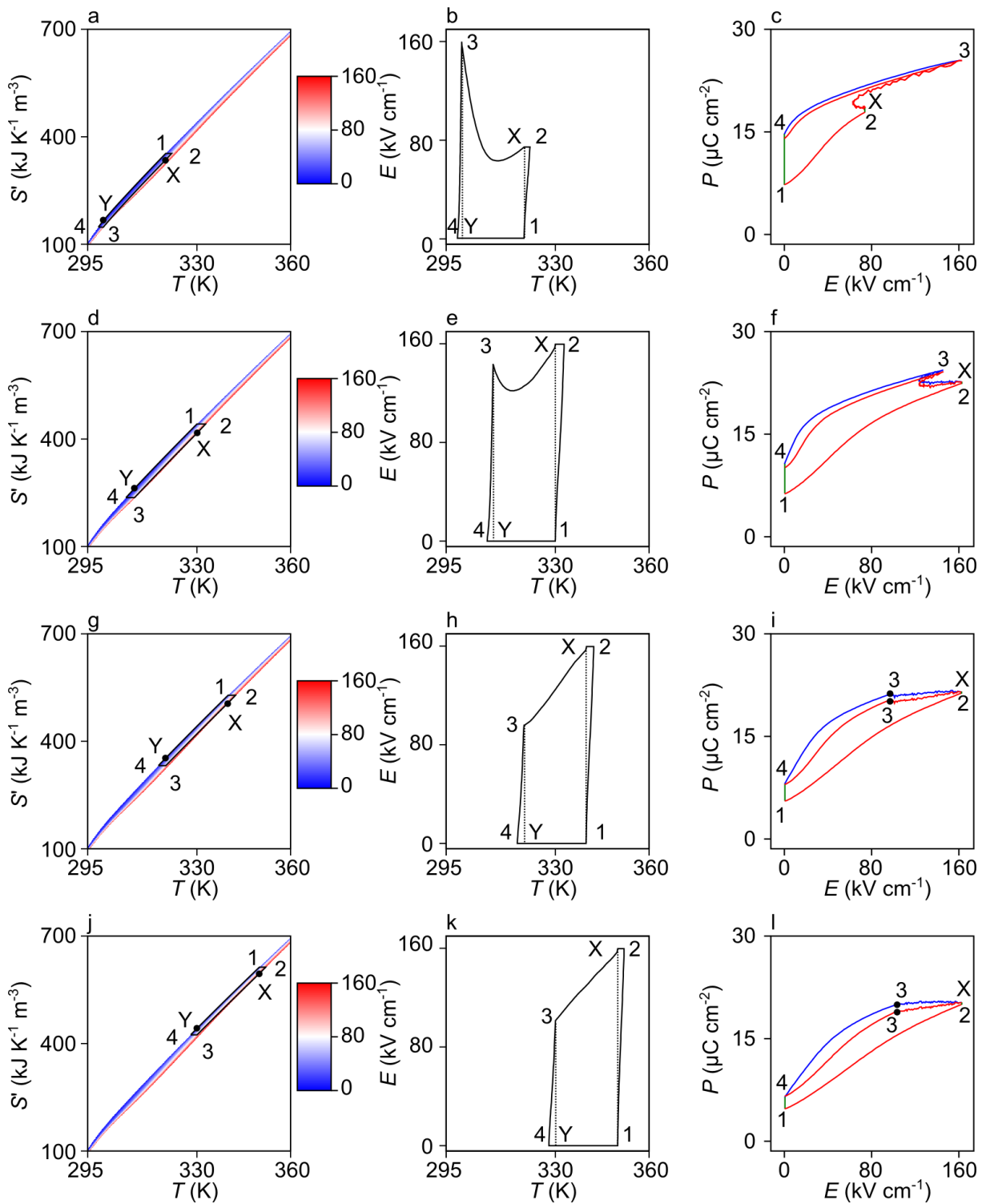

**Figure S16. Representative Brayton-like cycles with true regeneration for clamped PST above $T_C$.** The panels of Fig. S15a,b,f are redrawn for cycles with (a-c) low, (d-f) intermediate and (g-i) high values of $T_c$ and $T_h$. Each cycle reaches $E_{max} = 160$ kV cm$^{-1}$ during 2-X-3. Non-monotonicity in (b,c,e,f) evidences a small mismatch between absolute temperatures in the underlying electrical and thermal data.

**Note 15. Repeat of Fig. S15e,h with COP instead of efficiency**

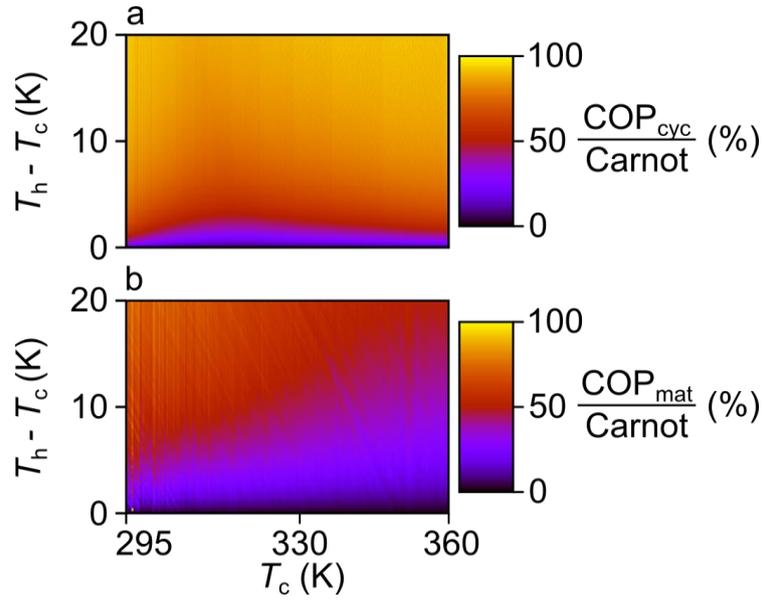

**Figure S17. COP$_{cyc}$ and COP$_{mat}$ as fractions of the Carnot limit for clamped PST.** (a) COP$_{cyc}$ = $Q/\oint T dS'$ (Fig. S15e) is presented here after dividing through by the Carnot limit $T_c/(T_h - T_c)$. (b) COP$_{mat}$ = $Q/\oint E dP$ (Fig. S15h) is presented here after dividing through by the Carnot limit $T_c/(T_h - T_c)$. Cross-sections through each panel at $T_h - T_c$ = 10 K and 20 K appear in Fig. 5d of the main paper.
52

## Note 16. Comparison of EC properties for PST and clamped PST

As seen in Table 1, PST outperforms other bulk EC materials in terms of $|\Delta T|$, $|\Delta S|$ and refrigerant capacity RC. The performance of bulk $PbSc_{0.5}Ta_{0.5}O_3$ studied elsewhere[S2] is similar, but here we identify a four-fold improvement in EC strength $|\Delta S|/|E|$ by recognising that $|\Delta S|$ saturates well below $E_{max}$. This EC strength is almost as large (~73%) as the value obtained in a narrow range of temperatures near the sharp 400 K transition in single-crystal $BaTiO_3$ (ref. S3), but here it remains large over a wide range of temperatures near room temperature.

As also seen in Table 1, clamped PST possesses larger values of $|\Delta T|$ and $|\Delta S|$ than bulk EC materials, because it was possible to apply a larger field without breakdown. However, the most notable improvement in performance is the increased temperature range of operation $T_2 - T_1 \sim 70$ K, and the resulting increase in refrigerant capacity RC ~ 152 J kg$^{-1}$, both of which exceed the values recorded for bulk EC materials by a factor of at least three. Clamped PST also displays a ~50% enhancement of EC strength $|\Delta S|/|E|$ with respect to ~1 μm-thick EC films, for which $E_{max}$ is disproportionately large.

|  | $|\Delta T|$ | $|\Delta S|$ | $|E|$ | $|\Delta S|/|E|$ | $T_2 - T_1$ | $T_C$ | RC | Ref. |
|---|---|---|---|---|---|---|---|---|
|  | K | J K$^{-1}$ kg$^{-1}$ | kV cm$^{-1}$ | J K$^{-1}$ kg$^{-1}$ kV$^{-1}$ cm | K | K | J kg$^{-1}$ |  |
| **Bulk samples** |  |  |  |  |  |  |  |  |
| PST | 2.3 | 2.6 | 26 | 0.4 | 17 | 295 | 43 | This work |
| $PbSc_{0.5}Ta_{0.5}O_3$ | 1.7 | 2.7 | 25 | 0.1 | 18.1 | 295 | 40 | S2 |
| $BaTiO_3$ | 0.9 | 2.2 | 4 | 0.6 | 1.5 | 397 | 3.2 | S3 |
| $BaTiO_3$ | 0.9 | 1.0 | 12 | 0.1 | 13.9 | 402 | 27.5 | S3 |
| PZST | 2.5 | 2.9 | 30 | 0.1 | 13.4 | 430 | 30.7 | S4 |
| **Thinned bulk** |  |  |  |  |  |  |  |  |
| clamped PST | 3.3 | 3.7 | 160 | 0.03 | 70 | 295 | 152 | This work |
| 80 μm PMN-PT | 3.5 | 3.2 | 160 | 0.02 | - | 400 | - | S5 |
| **Thin films** |  |  |  |  |  |  |  |  |
| $PbZr_{0.95}Ti_{0.05}O_3$ | 12 | 8 | 480 | 0.02 | 133 | 500 | 890 | S6 |
| PVDF-TrFE-CFE | 12 | 60 | 3000 | 0.02 | 17 | 330 | 814 | S7 |
| PST | 7 | 6.4 | 774 | 0.01 | 130 | 341 | 662 | S8 |

**Table 1. Figures of merit for EC materials near Curie temperature $T_C$.** Peak values of $|\Delta T|$, $|\Delta S|$ and $|\Delta S|/|E|$ due to field change $|E|$, where $|E| = E_{max}$. The exception arises for PST because $|\Delta S|/|E|$ is maximised at $|E| = 5$ kV cm$^{-1}$ < $E_{max}$ (Fig. 3c in the main paper). Refrigerant capacity RC = $\int_{T_1}^{T_2} |\Delta S(T)| dT$, where $T_1$ and $T_2$ define the FWHM of $|\Delta S(T)|$ for the removal of $E_{max}$ (Figs S6f, S14k). PZST = $Pb_{0.99}Nb_{0.02}(Zr_{0.75}Sn_{0.20}Ti_{0.05})_{0.98}O_3$, PMN-PT = $0.9PbMg_{1/3}Nb_{2/3}O_3$-$0.1PbTiO_3$, PVDF-TrFE-CFE = Poly(vinyledene flouride-triflouroethylene-chlorofluoroethylene).

**Supplementary references**